\documentclass[lettersize,journal]{IEEEtran}
\usepackage{amsmath,amsfonts}
\usepackage{algorithmic}
\usepackage{algorithm}
\usepackage{array}
\usepackage[caption=false,font=normalsize,labelfont=sf,textfont=sf]{subfig}
\usepackage{textcomp}
\usepackage{stfloats}
\usepackage{hyperref}
\usepackage{verbatim}
\usepackage{graphicx}
\usepackage{cite}
\usepackage{bm}
\usepackage{multirow}
\usepackage[font=footnotesize,justification=centering,compatibility=false]{caption}
\usepackage{xurl}
\usepackage{url}
\usepackage{float}
\usepackage{xcolor}
\usepackage{cancel}
\usepackage[normalem]{ulem}
\usepackage{wrapfig}

\usepackage{comment}
\usepackage{threeparttable}
\usepackage{comment}
\usepackage{tabularx}
\usepackage{adjustbox}

\begin{document}

\title{CMM-MP: A Novel Memory Pooling Architecture using Purpose-built Memory Device with CXL and Ethernet}

\author{
      
      Michael Wasef, %\authotnotemark{<m.wasef@samsung.com>}
      Gongjin Sun, % <gongjin.s@samsung.com>
      Edmund Au,  %\authotnotemark{<edmund.au@samsung.com>;} 
      Ranjitha Gopalakrishna,  %<ranjitha1.g@samsung.com>; 
      Elham Movahedian, % <elham.m@samsung.com>; 
      Dongyang Li, % <dongyang.li@samsung.com>
      Senthil Murugesapandian, % <s.murugesapa@samsung.com>; 
      Caroline Kahn, % <c.kahn@samsung.com>;
      Shuyu Lyu, Brian Luu, Jason Scott, Namita Kumar,
      Mehdi Nik, %<mehdi.Nik@samsung.com>
      Kwan Huen,
      Xuebin Yao, % <xuebin.yao@samsung.com>; 
      Amir Beygi, % <amyr.beygi@sasmung.com>
      Luis Vitorio Cargnini, \IEEEmembership{Member, IEEE}, % <v.cargnini@sasmung.com>
      Ramdas Kachare%<r.kachare@samsung.com>
      %\thanks{Manuscript received November 4, 2023; revised February 28, 2024.}
% \thanks{The next few paragraphs should contain 
% the authors' current affiliations, including current address and e-mail. For 
% example, First A. Author is with the National Institute of Standards and 
% Technology, Boulder, CO 80305 USA (e-mail: author@boulder.nist.gov). }
\thanks{This work has been submitted to the IEEE for possible publication. Copyright may be transferred without notice, after which this version may no longer be accessible.

All authors belong to Samsung Semiconductor Inc., Memory Solutions Laboratory -- Data Center Device Solutions Group, San Jose, CA, 95134 USA. (e-mail(s):\{r.kachare, mehdi.nik, m.wasef, gongjin.s\}@samsung.com).}
 }

%\markboth{IEEE Journal of \LaTeX\ Class,~Vol.~12, No.~6, February~2024}%

%\IEEEpubid{0000-0000~\copyright~2024 IEEE}

\maketitle

\begin{abstract}

\looseness-1 CXL Memory Module – Memory Pooling (CMM-MP) introduces a novel memory pooling architecture to meet the unique requirements of emerging memory-intensive workloads. This industry-first, purpose-built memory-pooling device combines the benefits of Compute Express Link (CXL) with the scalability and cost advantages of Ethernet. Multiple CMM-MP devices together provide a flexible, composable, and disaggregated memory expansion solution. The CMM-MP enables 100s of TB capacity for GPU/CPU applications without modifications. The memory pool is exposed as a large-capacity logical aggregate memory, with all pool discovery, composition, and backend network datapath tasks fully offloaded to the device. CMM-MP supports both memory pooling and sharing usage modes. It provides an application latency of 2.7 us and up to 23 GB/s of memory throughput. Our evaluations demonstrate that CMM-MP significantly outperforms RDMA-based solutions in latency and throughput. We also evaluated real-world use cases of Valkey+YCSB and DiskANN using CMM-MP.        

\end{abstract}

\section{Introduction}

\IEEEPARstart{T}{he} rapid growth of memory-intensive applications in modern datacenters, including machine learning, graph processing, cloud-native services, and large language models, has intensified the demand for large, adaptable memory capacity\cite{ref43}. Conventional server architectures limit the amount of memory available to the running applications. The current system architectures lead to underutilization and memory scarcity, depending on the application workload\cite{ref44}. As workloads evolve, it is difficult to predict the proper memory capacity requirements in advance. Power, space, and thermal constraints limit the ability to scale memory independently\cite{ref45, ref46}. As a result, applications whose peak memory requirements exceed the limits of a single node are poorly served by the traditional tightly integrated memory model. Another challenge for data center operators is efficiently sharing idle or lightly loaded server memory with other hosts without compromising performance, thereby impacting the overall TCO of data centers\cite{ref30}. The core challenge is to devise a scalable memory‑expansion architecture that meets datacenter constraints of TCO, performance, power, and cooling\cite{ref23}. This leads to a more fundamental realization that the current ecosystem is heavily compute-centric, whereas emerging workloads require in-depth memory architectures and data-loading considerations.

Memory disaggregation\cite{ref12, ref37, ref38} alleviates these constraints by decoupling memory from individual compute nodes and exposing it as a shared accessible resource over high‑bandwidth, low‑latency interconnect\cite{ref49, ref23}. This separation enables dynamic allocation, reduces over‑provisioning, improves utilization, and allows data center operators to scale compute and memory independently. By breaking the rigid linkage between compute and memory, disaggregation provides the necessary elasticity to support data‑intensive workloads\cite{ref39}.  

Building on the flexibility of memory disaggregation, memory pooling consolidates multiple memory devices into a unified, addressable resource shared across many compute nodes. By leveraging a common memory pool, the system minimizes under-utilization and dynamically adapts to workload variations. This pooled‑memory approach improves scalability by avoiding over‑provisioning of host system memory, thereby reducing Total Cost of Ownership (TCO) and power consumption\cite{ref30}. Memory pooling can also improve performance and reduce power consumption by reducing data movement. Use cases for in-place memory sharing could include virtual machine (VM) migrations and multiple hosts sharing data\cite{ref41, ref50}.   

\begin{figure}[h!]
    \centering
    \includegraphics[width=0.95\columnwidth,page=29]{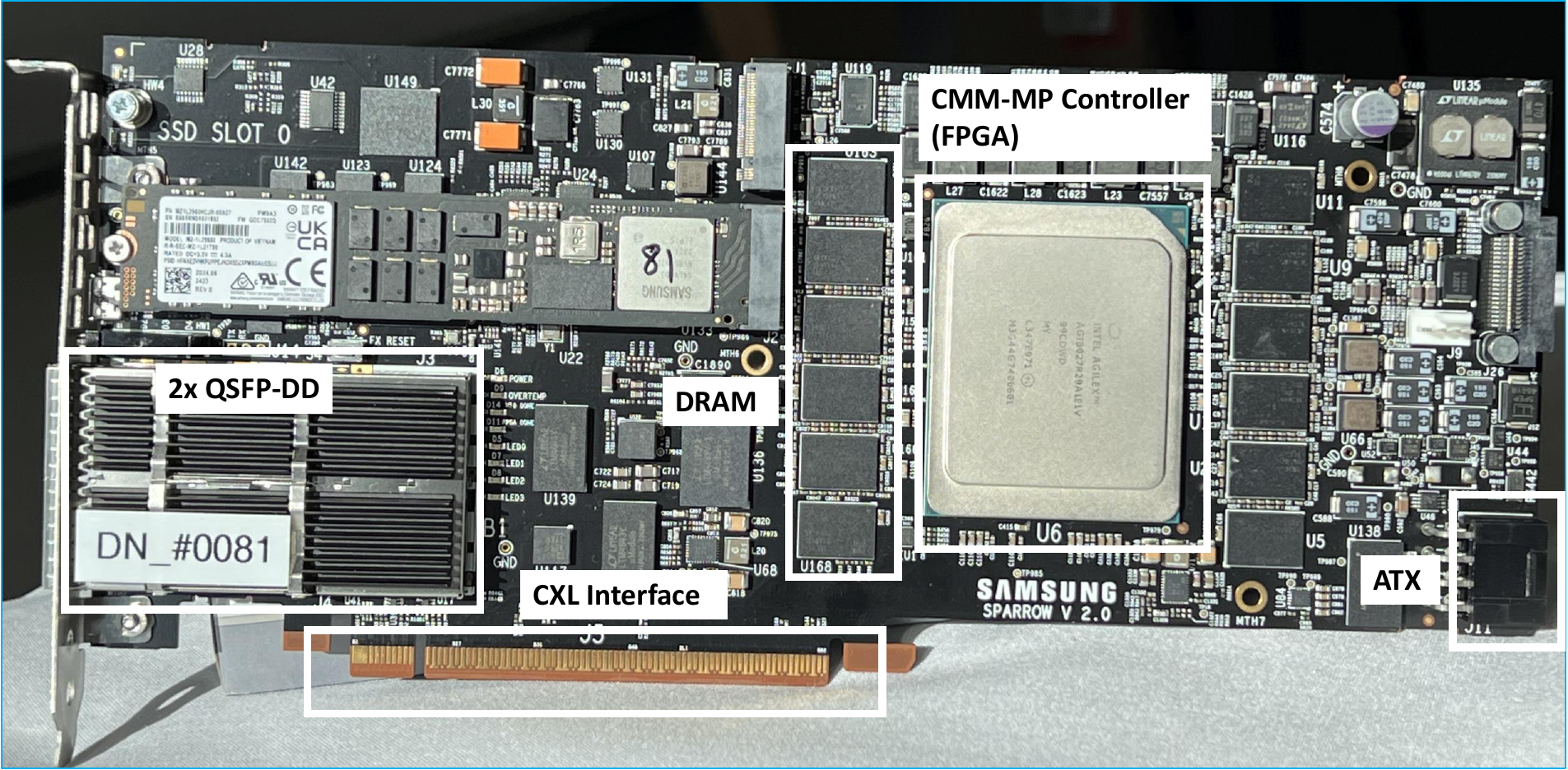}
    \caption{CMM-MP device}
    \label{fig:29}  
\end{figure}

\begin{figure}[h!]
    \centering
    \includegraphics[width=0.95\columnwidth,page=1]{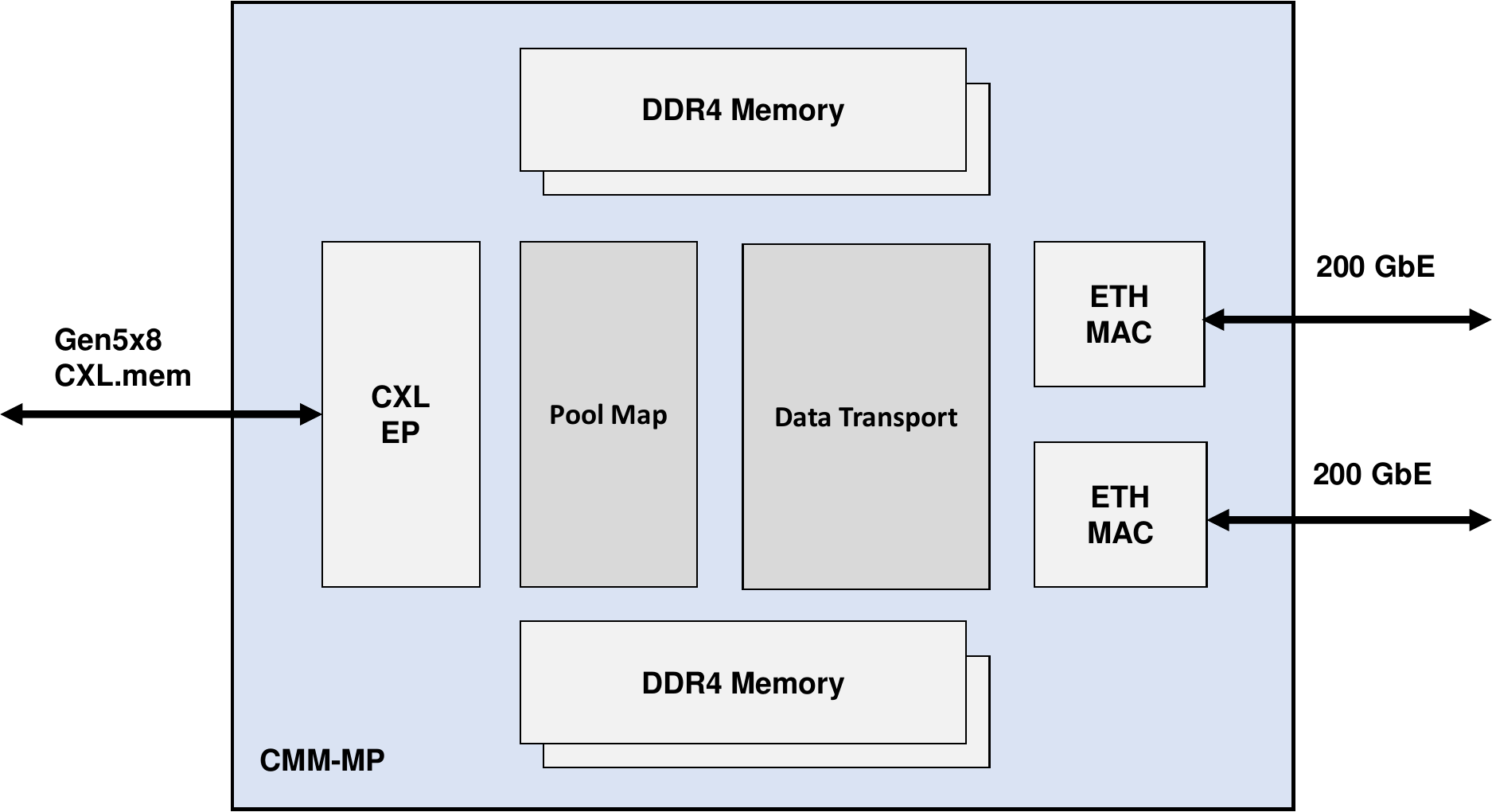}
    \caption{CMM-MP device block architecture}
    \label{fig:1}  
\end{figure}

This paper introduces CMM-MP, a novel memory pooling device and system architecture that addresses some of the challenges mentioned above. As shown in Figure \ref{fig:29}, this purpose-built CMM-MP device combines the benefits of CXL host interface with the scalability and cost advantages of the Ethernet backend. The CMM-MP architecture utilizes CXL's memory (load/store) semantics. This avoids any changes to the application to utilize the memory pool. In the CMM-MP device, Ethernet serves as the transport fabric forming a memory pool. Figure \ref{fig:1} shows a high-level device architecture comprising a CXL Endpoint host interface, local DRAM memory, memory pooling logic, and network ports. Memory pooling logic comprises memory address translation, memory pool composition, and reliable data transport over the backend network interface.  The entire memory pool is exposed to its local host as a logical system memory address range. 

\begin{figure}[h!]
    \centering
    \includegraphics[width=\columnwidth,page=2]{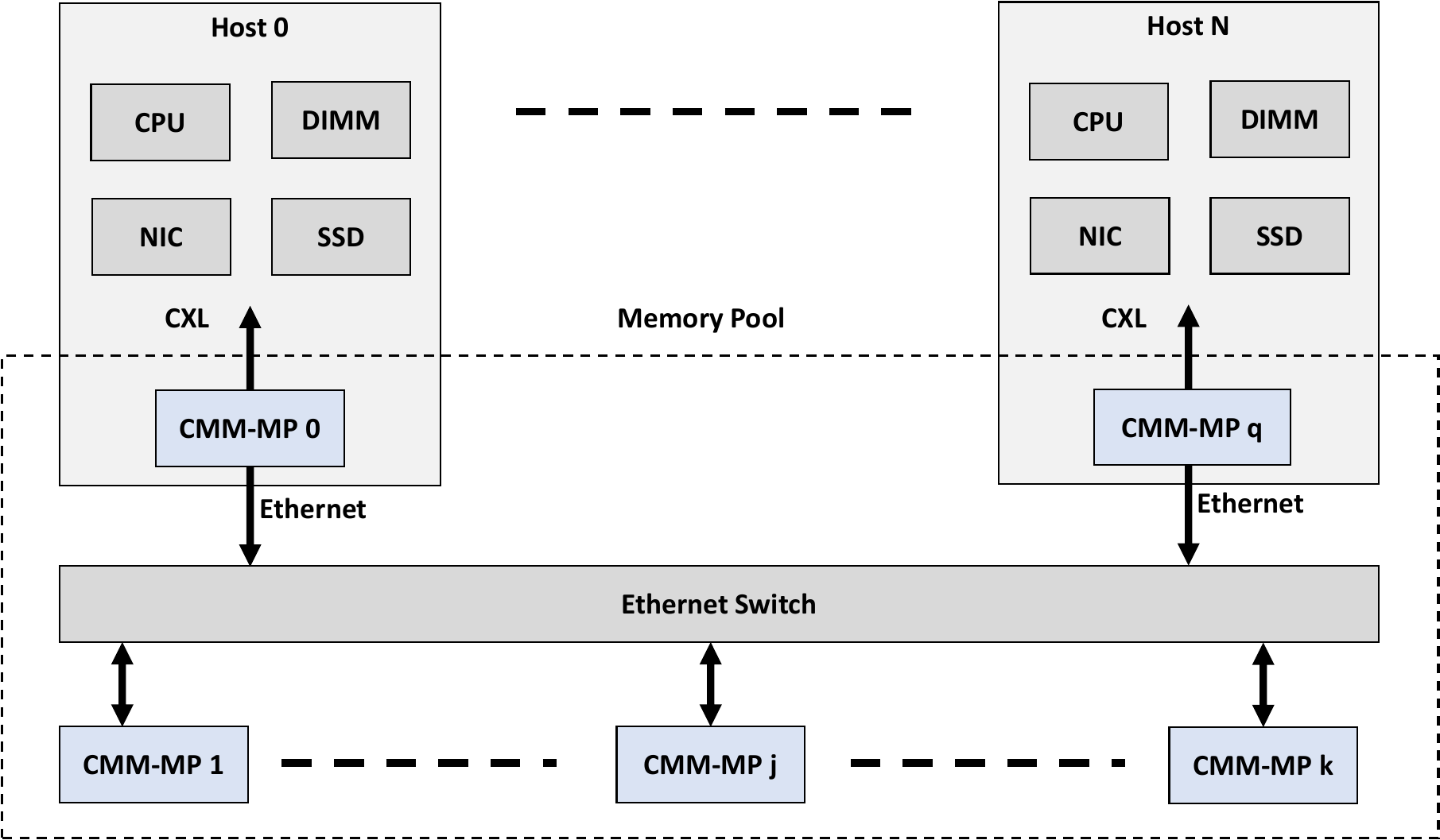}
    \caption{CMM-MP memory pooling system architecture}
    \label{fig:2}  
\end{figure}

\begin{figure}[h!]
    \centering
    \includegraphics[width=\columnwidth,page=3]{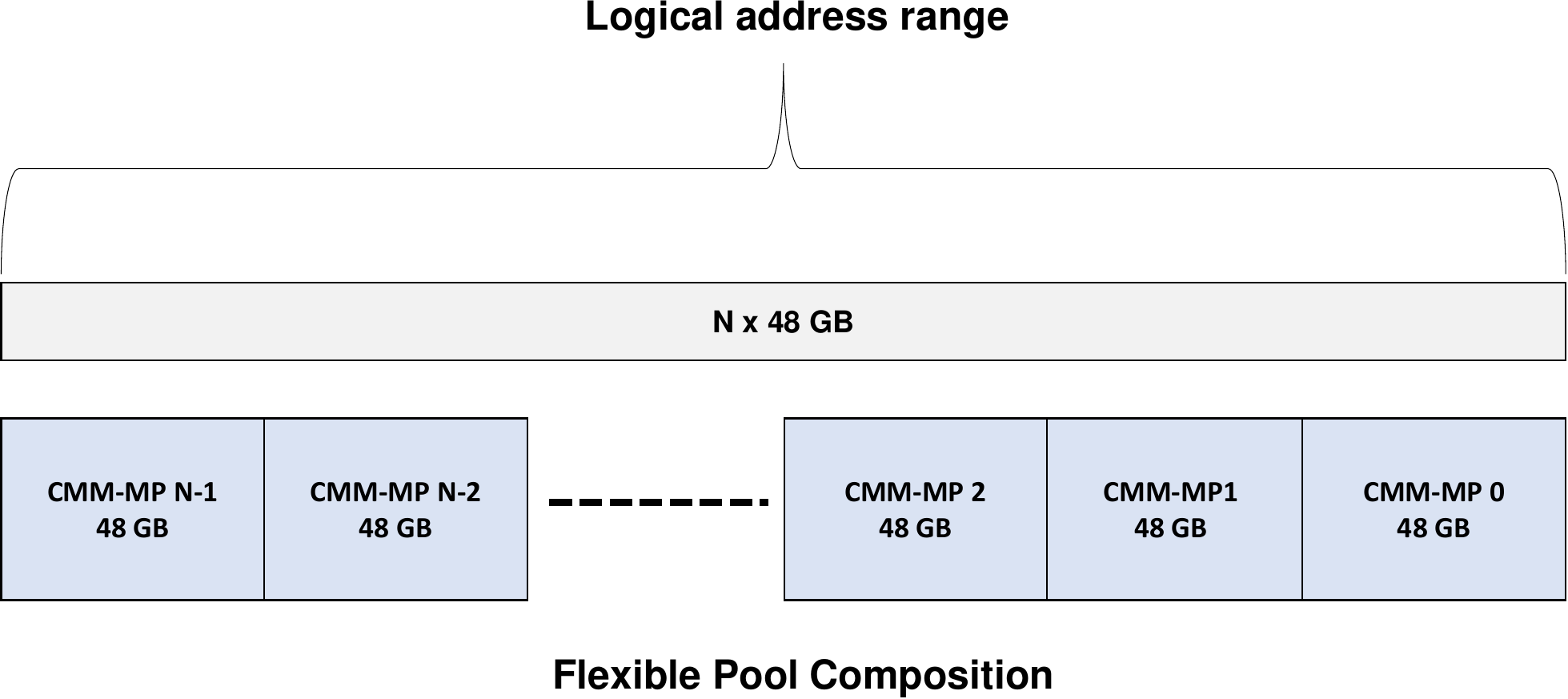}
    \caption{Memory pool hierarchy}
    \label{fig:3}  
\end{figure}

As illustrated in Figure \ref{fig:2}, the memory pool consists of multiple CMM-MP devices connected through a dedicated Ethernet switch. Each participating host contains one local CMM-MP device. All the CMM-MP devices are connected using a dedicated backend Ethernet switch. The CMM-MP device firmware autonomously discovers available pool devices and their memory capacity without burdening any host. Each host can access the whole pool capacity through its local CMM-MP device. The pool capacity seen by each host is composed as shown in Figure \ref{fig:3}. CMM-MP host management software utility manages pool memory allocation policy.   

The paper is organized as follows: we first review related work on memory disaggregation and pooling, then describe the CMM-MP architecture. Next, we explain the CMM-MP discovery protocol and the process of creating a memory pool. Afterward, we detail the host interface for memory allocation management. This is followed by a comprehensive performance evaluation of latency and bandwidth. Further, we test multiple real-world applications using CMM-MP. Finally, we conclude with a summary of this work and directions for future research.

\section{Background and related work}

Memory disaggregation has been pursued through a variety of technologies, including RDMA, CXL, and proprietary technologies. 

A common technique for memory pooling is Remote Direct Memory Access (RDMA)\cite{ref17}. RDMA can provide scale and capacity for memory resources. However, it incurs the overhead of explicit remote‑memory calls and intrusive software changes\cite{ref47, ref48}.
The RDMA approach has improved latency and throughput in key-value systems \cite {ref19}. Aggregating DRAM from multiple servers into a shared pool accessible via RDMA is shown to improve the performance of remote memory\cite{ref20}.

As for the CXL solutions\cite{ref23}, they can overcome the software complexity of RDMA by offering native load/store semantics\cite{ref50} seamlessly to unmodified applications\cite{ref32}, remaining confined to rack‑level deployments.  Memory pooling through a dedicated CXL switch is another solution\cite{ref29, ref42} that enables memory expansion and improves coherence, but it has limited scalability, higher cost, and deployment constraints. Efforts have been made to extend the CXL switch architecture with an external memory controller\cite{ref30} to meet cloud performance requirements and reduce memory costs. 

Another solution introduced for CXL2.0 uses the Multi-Logical Device (MLD)\cite{ref40} feature with the CXL switch. It provides transparent, cache-line-granular memory sharing across hosts. However, this approach is limited by physical reach (rack-scale only), switching latency, higher costs, and scalability challenges. The CXL MLD concept later evolved into a Multi-Headed Device in CXL 3.0\cite{ref41}, effectively eliminating the physical CXL switch. It enables a single memory device to expose multiple memory channels to the shared pool, but the physical reach limitation remains.

Hybrid designs that combine CXL and RDMA\cite{ref18} aim to capture the best of both worlds, yet they still inherit RDMA’s overheads and the complexity of an additional software layer on the host, preventing a seamless, transparent memory‑pooling solution.

Another CXL memory approach is CXL Memory Module - Hybrid (CMM-H). It  combines DRAM and NAND memory to provide a low-cost, large capacity memory expansion solution\cite{ref35}\textsuperscript{,}\cite{ref36}
. This approach provides a large memory capacity, but it does not support the memory sharing feature.  

CXL‑over‑Ethernet \cite{ref33} is yet another approach that preserves the low‑latency benefits of CXL for individual remote accesses but does not provide a genuine pooled memory accessible by multiple servers.

Other approaches have attempted to reuse existing interconnects. For example, connecting two CPUs via the Ultra Path Interconnect (UPI)\cite{ref52} and Ethernet \cite{ref53} enables memory sharing within a single server but is limited in scalability, short-range, and consumes high power.  

\section{Device Architecture}

\begin{figure}[!h]
    \centering
    \includegraphics[width=\columnwidth,page=4]{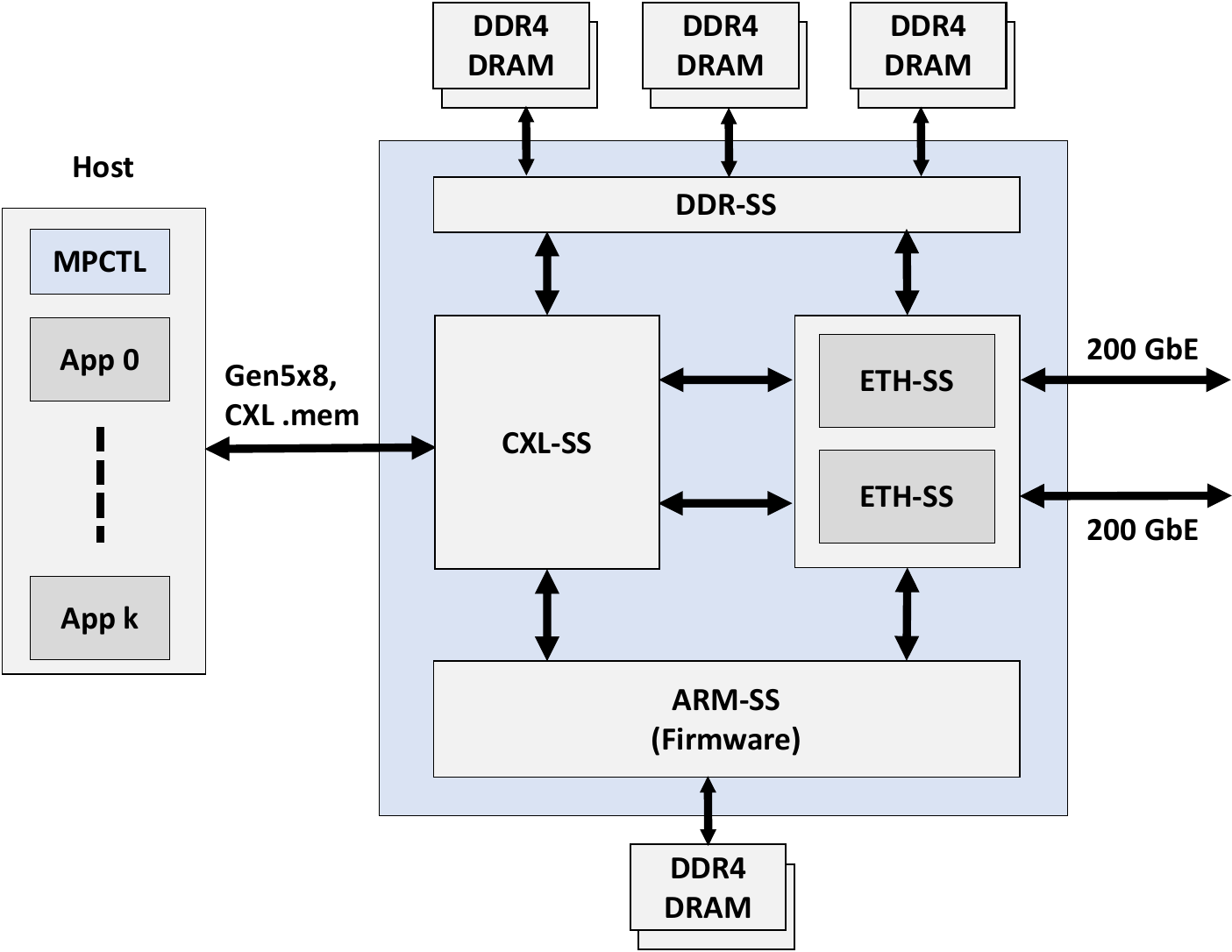}
    \caption{Memory pooling device micro architecture}
    \label{fig:4}  
\end{figure}

CMM-MP is a purpose-built device to implement a high-performance memory pooling solution. It is a CXL Type 3 memory device that integrates CXL and Ethernet in a novel architecture to provide large-capacity memory expansion. It contains four dedicated logic subsystems, as shown in Figure \ref{fig:4}. 

The CXL Subsystem (SS) implements Gen5x8 CXL .mem protocol to process host memory read/write requests. This subsystem determines whether each host request targets local/remote memory and forwards them to the corresponding subsystems (DDR or Ethernet) for data access. 

The Ethernet Subsystem includes two 200 Gbps Ethernet ports (total 400 Gbps of raw bandwidth) for backend connectivity between CMM-MP devices. This subsystem is responsible for packet processing, routing, and network reliability functions, including detection and mitigation of packet loss, duplication, reordering, and incorrect addressing. 

The DDR subsystem implements the physical memory capacity with DRAM packages and associated memory controllers. This subsystem arbitrates local requests from the CXL subsystem and remote requests from the Ethernet subsystem. 

The ARM subsystem contains an ARM Cortex A53 CPU to handle control-plane functions such as pool device discovery, pool memory-mapping composition, and programming the internal mapping table. The ARM subsystem concurrently injects and receives control packets into and out of the data path alongside host traffic.  

\begin{figure*}[!h]
    \centering
    \includegraphics[width=\linewidth,page=5]{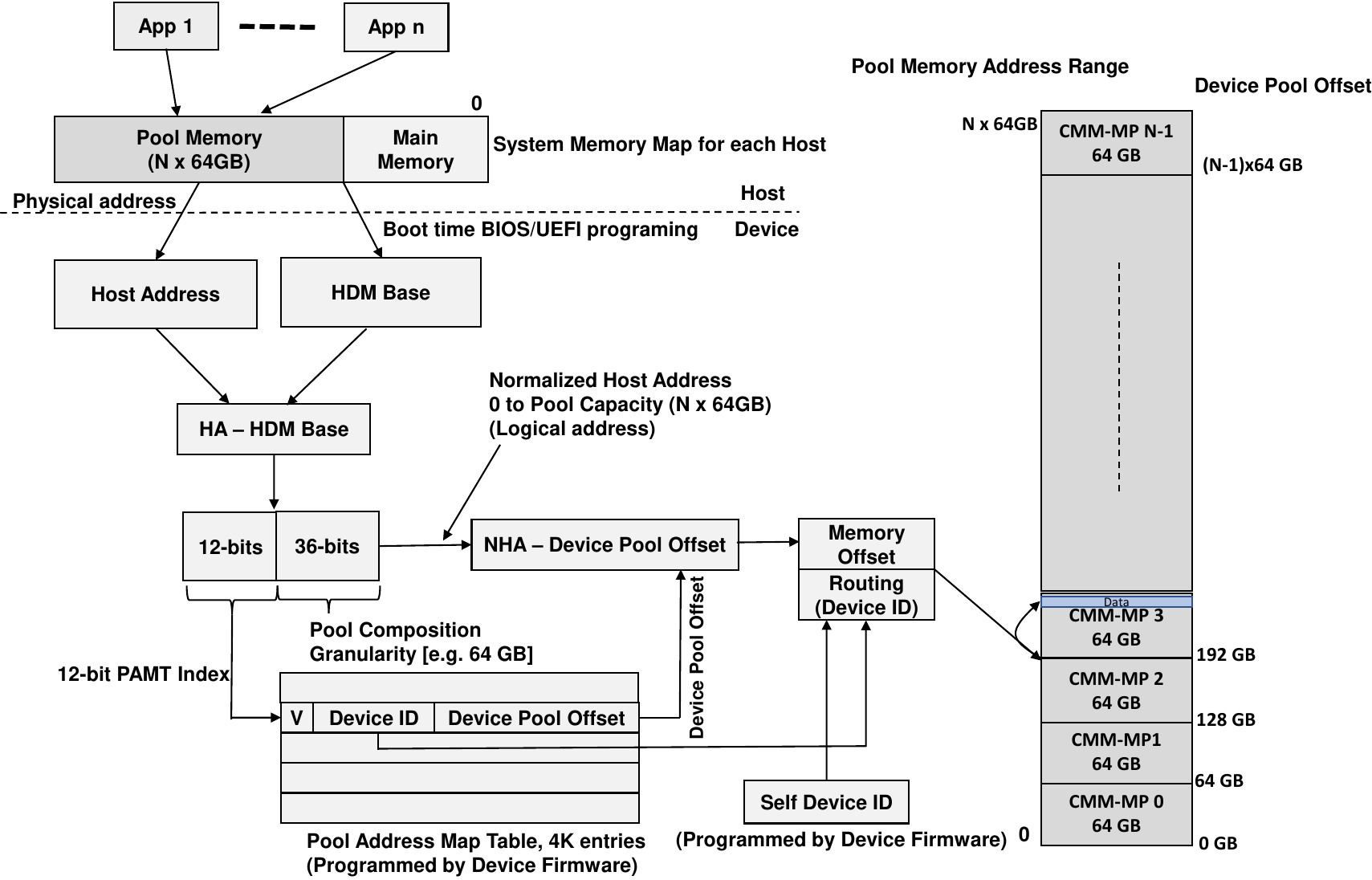}
    \caption{Address mapping and resolution in CMM-MP device}
    \label{fig:5}  
\end{figure*}
The remainder of this section provides the host request processing flow and details of each subsystem.

\subsection{Memory Pool Address Resolution}
The CMM-MP device performs a critical function: mapping and translating host addresses to appropriate internal pool memory addresses. This allows participating hosts to share a common, consistent view of the pool memory address range, which enables seamless access. This process is shown in Figure \ref{fig:5}. 
Each CMM-MP device on a server node advertises the entire pool capacity to the host. During server boot, BIOS/UEFI maps the advertised pool capacity to the system memory map. The physical base address of the mapped pool memory is programmed to the Host Managed Device Memory (HDM) Base Address register of the device. During runtime, when applications access the allocated pool memory, those memory requests are fulfilled via the CMM-MP device. The Host address (HA) range of these memory requests is from HDM Base to {HDM Base + Pool Capacity}. The Normalized Host address (NHA), that spans from zero to the pool capacity (logical address), is determined as follows:
\begin{equation}
NHA = HA - HDM \ Base 
\end{equation}

The next step is to locate the data corresponding to this normalized host address. This is achieved by looking up NHA in the Pool Address Map Table (PAMT). PAMT contains 4K entries of configurable granularity (e.g., 1GB, 4GB, ... up to 64GB). Using 64 GB granularity as an example, the pool can reach a maximum capacity of 256 TB. As shown in Figure \ref{fig:5}, a 12-bit index is used to look up the PAMT entry corresponding to the normalized host address. The valid PAMT entry provides two pieces of information: 1) Device ID, and 2) Device Pool Offset. The PAMT Device ID is used to route traffic to the target device. The Device Pool Offset is used to determine the memory offset within the target device memory. 
\begin{equation}
Memory~Offset = NHA - (Device~Pool~Offset)
\end{equation}
The target CMM-MP device uses the memory offset to access the requested data and then sends the response back to the requesting CMM-MP device. When the PAMT Device ID matches the self Device ID, data is accessed locally without going over the Ethernet network.

\subsection{CXL Subsystem}
\begin{figure}[!hb]
    \centering
    \includegraphics[width=\columnwidth,page=6]{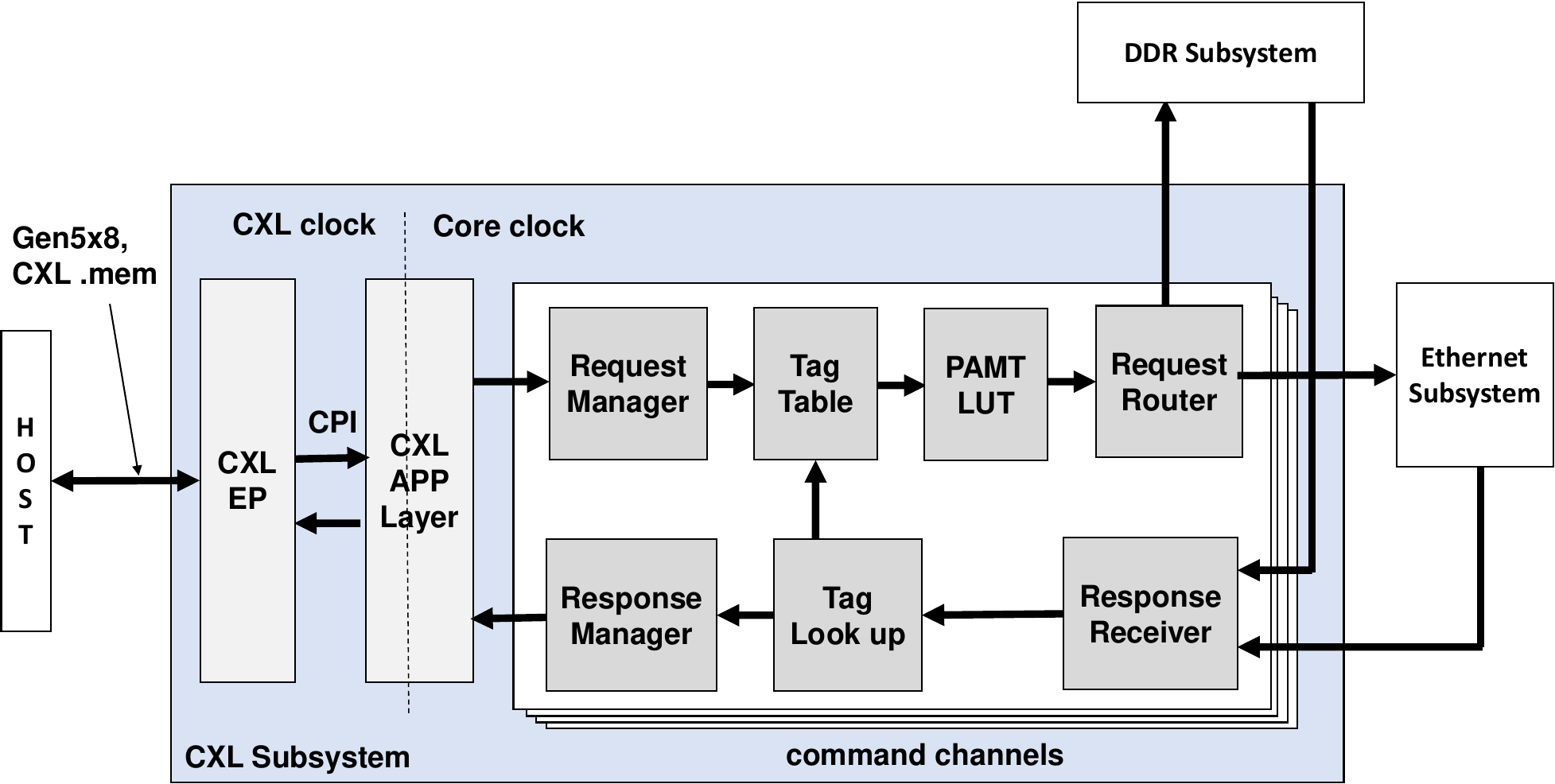}
    \caption{CXL subsystem architecture}
    \label{fig:6}  
\end{figure}

The CXL subsystem, as shown in Figure \ref{fig:6}, serves as an interface between the host and the device. It implements the CXL specifications $1.1$/$2.0$ (partially) over a PCIe Gen5 x8 link. It interprets incoming CXL read (MemRd) and write (MemWr, MemWrPtl) transactions. It also supports memory invalidation (MemInv) and speculative memory access (MemSpecRd). This subsystem maps the received host addresses to pool memory addresses.

The CXL Endpoint (EP) module implements the PCIe Link Training and Status State Machine (LTSSM). It implements PCIe/CXL Transaction, Data Link, and Physical Layer functions as specified in the PCIe/CXL standards. This module extracts CXL requests in the CXL clock domain and provides them via Coherent Processing Interface (CPI) for further processing.  

The CXL Application (CAPP) module performs clock-domain crossing (CDC) between the CXL clock and the internal core clocks. It implements CXL flow‑control by managing credit allocation and back-pressure signaling to the CXL-EP module. This module initializes CXL credits during boot and replenishes them at runtime. Incoming host CXL requests are distributed across four channels for concurrent processing. This module receives fully formed host responses, No Data Response (NDR), and Data Response (DRS) from the Response Manager. The NDR and DRS messages are returned to the host via the CXL EP module. It performs a similar CDC function on the outgoing response path.      

The CXL Request Manager parses the received CXL requests to extract header and payload fields. This module maps the host address in the CXL request to the logical, aggregate pool memory address (called Normalized Host Address, NHA) using equation 1. This address normalization step is critical for multiple hosts to have a consistent view of the pooled memory, since each participating host maps the pooled memory at different locations in its own system memory map.   

The Tag Table module manages metadata for in-flight requests throughout their lifecycle. For each CXL request, the Tag Table allocates a free slot and stores the header fields needed to create CXL responses later. The stored header fields include the memory operation code (MemOpcode), MetaField, MetaValue, Logical Device Identifier (LD‑ID), and the CXL request tag. The index of the allocated Tag Table is used as an internal tag to identify each request. It is carried through the datapath pipeline along with necessary request fields such as address and MemOpcode. 

The Pool Address Map Table (PAMT) stores the composition of pooled memory, as shown in \ref{fig:5}. Each CMM-MP device in the pool is assigned a unique device identifier (Device ID). The PAMT Lookup Table (LUT) module translates an NHA into the corresponding target CMM-MP Device ID. This operation also provides the offset within the target CMM-MP device's memory, thereby resolving the requested data location.  

The PAMT entry contains a valid bit, a device ID, and the Device Pool Offset. The device ID field is used to route the request to the proper target device in the pool. The Device Pool Offset field specifies the memory offset within the target device, as shown in figure \ref{fig:5}. The PAMT LUT module forwards the computed memory offset and device ID to the Request Router module for further processing.   
The Request Router module first compares the PAMT Device ID with the self Device ID. If the PAMT device ID matches the self device ID, the data request is fulfilled locally. Otherwise, it forwards the PAMT device ID, memory offset, internal Tag index, and other associated request information to the Ethernet subsystem for remote data access. 

On the data response side, the Response Receiver module collects completed responses from either the DDR or the Ethernet subsystems. Every response includes the internal Tag, along with any applicable data. The Tag Lookup module uses the internal Tag to get the original CXL Request header fields. Once the relevant header fields are read, the Tag table entry is freed and can be allocated to new host requests. The Response Manager module uses the retrieved header fields and, if applicable, the received payload data to generate an appropriate host CXL response. Response Manager generates two types of CXL responses: NDR and DRS. For read requests, it assembles a DRS that includes the requested data payload. For write and memory-invalidation requests, it creates an NDR that contains status and completion codes. It then formats the responses according to the CXL transaction layer packet specification and forwards them to the CXL application layer for transmission to the host via the CXL EP.
\begin{figure}[!h]
    \centering
    \includegraphics[width=\columnwidth,page=10]{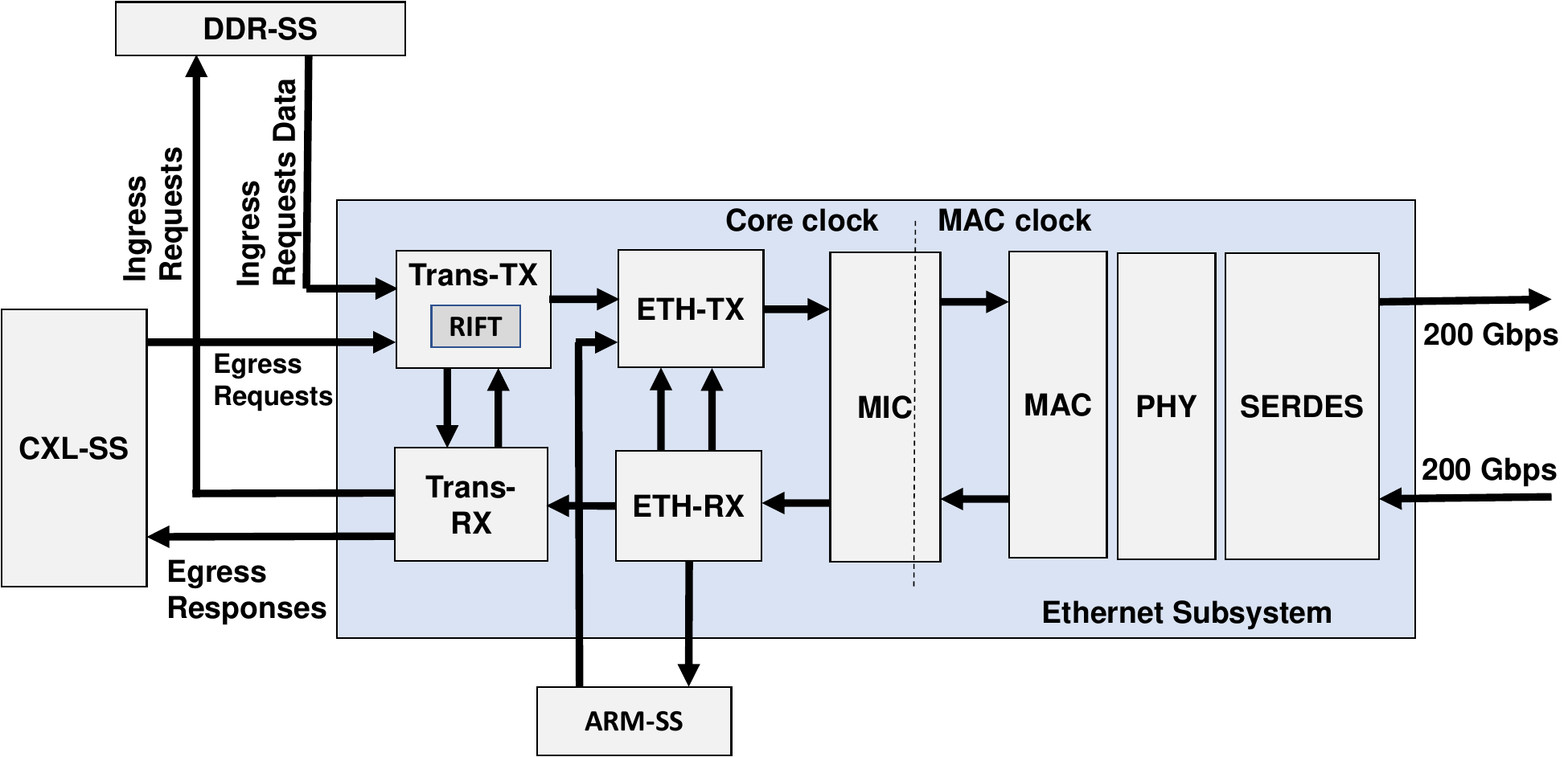}
    \caption{Ethernet subsystem micro architecture}
    \label{fig:7}  
\end{figure}

\subsection{Ethernet Subsystem}
The Ethernet subsystem provides a data transport service for memory pool operation. It implements three logical layers: the Transport Layer, the Data Link Layer, and the Physical Layer. It enables a host to access data on remote devices in the memory pool. Additionally, ARM firmware uses this subsystem to exchange control packets among the pool memory devices. It implements essential network reliability functions to provide lossless, high-throughput, low-latency data transport. The Ethernet subsystem ensures end‑to‑end reliability through automatic retransmission, duplicate-packet handling, and packet-loss detection and recovery. This provides robust and predictable data movement across the memory pool fabric. Figure \ref{fig:7} shows the micro-architecture of the Ethernet subsystem.

\begin{figure}[!h]
    \centering
    \includegraphics[width=\columnwidth,page=12]{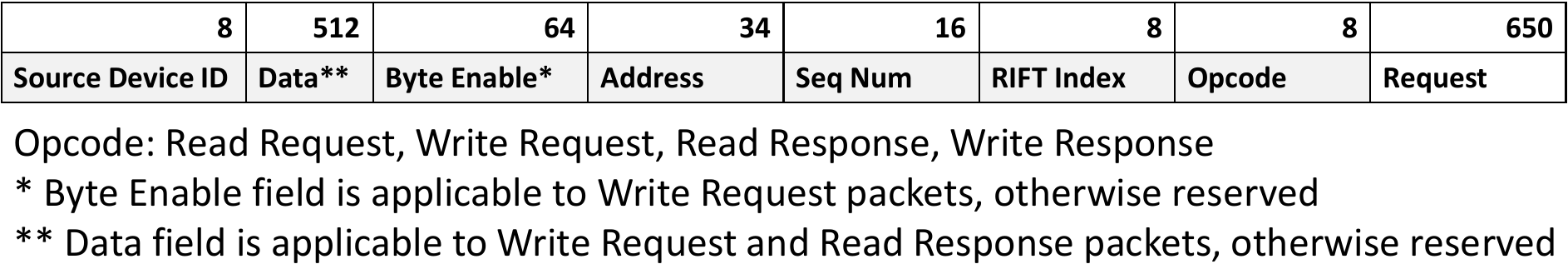}
    \caption{Transport packet format}
    \label{fig:8}  
\end{figure}

The Trans-TX module creates transport-layer packets for: 1) the egress remote memory requests, 2) the responses for the ingress requests. The constructed transport packets are forwarded to the Ether TX module for further processing. The composition of transport packets is shown in Figure \ref{fig:8}. The Egress remote memory request consists of the remote device ID, memory offset, and operation type (read/write). For every outstanding host remote memory request, the Trans-TX module creates and maintains a context data structure in the Request‑In‑Flight Table (RIFT).

\begin{figure}[!h]
    \centering
    \includegraphics[width=\columnwidth,page=11]{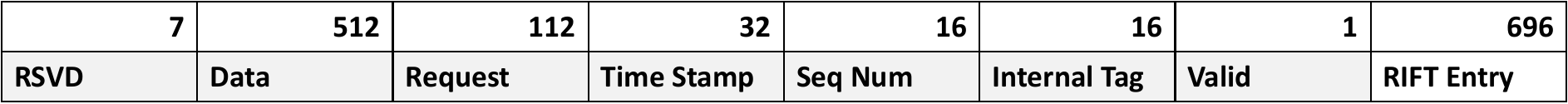}
    \caption{RIFT table entry data structure}
    \label{fig:9}  
\end{figure}

The format of a RIFT entry is shown in Figure \ref{fig:9}. The valid flag indicates an outstanding request; once the request is completed, the valid flag is reset, and the RIFT entry is freed for new requests. The Trans-TX module assigns a unique, monotonically increasing 16-bit transmit Sequence Number (SN) to each request. These sequence numbers are separate and independent for each remote device. The sequence number is used to detect lost and duplicate packets on the remote device. If the timestamp exceeds a threshold without a response, Trans-TX retransmits the request to ensure reliable operation of the data fabric. For ingress requests, this module constructs response packets using remote request context information from the Trans-Rx module and the associated data received from the DDR subsystem. 

The Trans-RX module processes two types of packets: 1) ingress requests, 2) data responses for egress requests. The opcode field, as shown in Figure \ref{fig:8}, in the received transport packet, is used to determine the packet type: request or response. To process ingress requests, the Trans-RX module maintains a receive sequence number table for all devices in the pool. This table serves as the foundation for handling out-of-order, duplicate, and lost packets. For each source device ID, the table stores a base sequence number. The base sequence number represents the highest in-order packet that has been successfully received. The table also contains a 64‑bit packet bitmap and an overflow counter for each source-Device ID. The packet bitmap is interpreted as shown in the equation below. 
\begin{equation}
  bitmap[i] =
    \begin{cases}
      1&\!\!\!\!\!base~SN+i+1 \text{~packet received}\\
      0&\!\!\!\!\!base~SN+i+1 \text{~packet expected}\\
    \end{cases}       
\end{equation}

\begin{figure}[!h]
    \centering
    \includegraphics[width=\columnwidth,page=13]{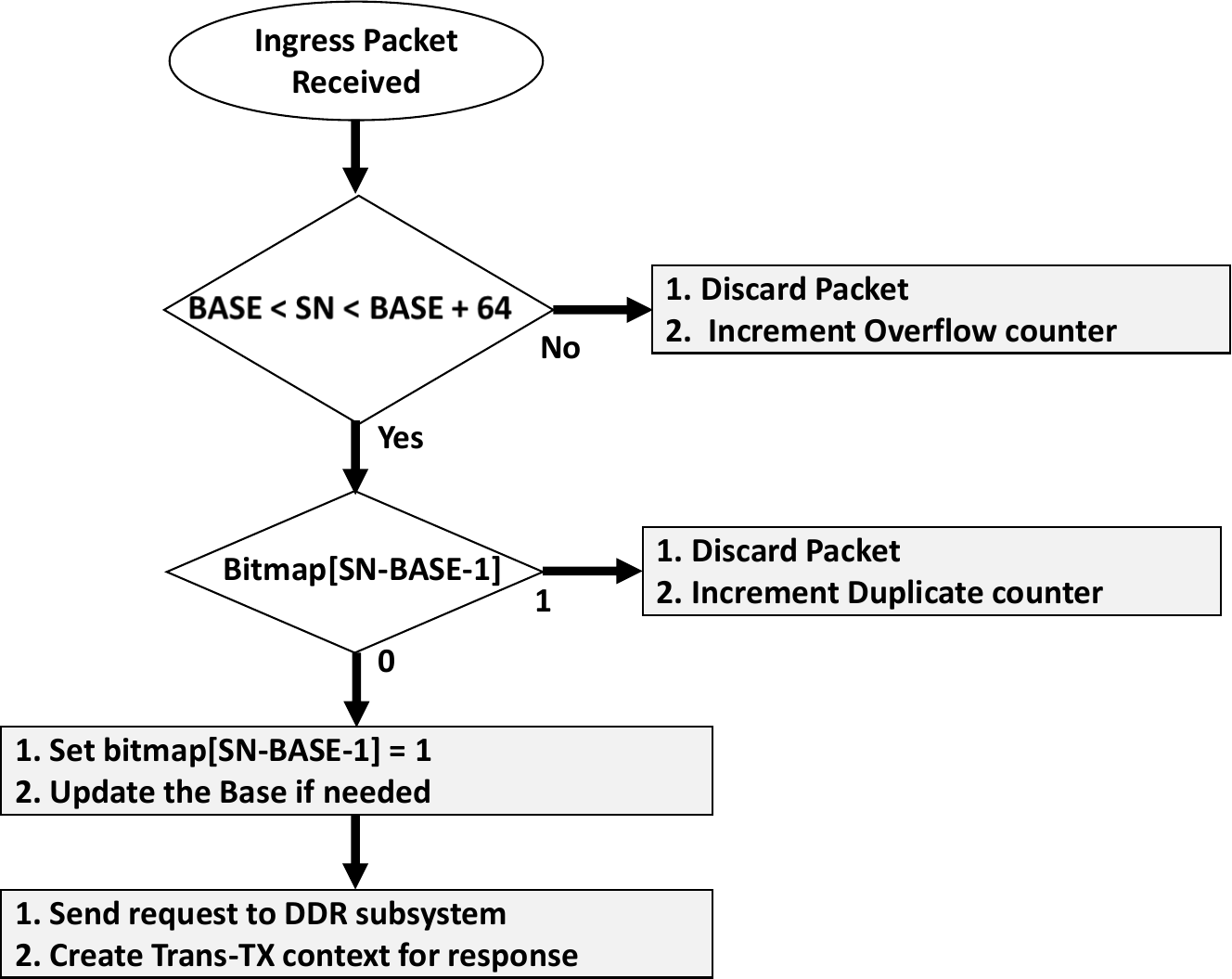}
    \caption{Algorithm for transport packet processing}
    \label{fig:10}  
\end{figure}

This representation enables detection and handling of out‑of‑order arrivals and duplicate requests. The overflow counter records how many packets fall beyond $base+64$ and triggers resynchronization when necessary. The ingress packet processing flow is shown in Figure \ref{fig:10}.

The base SN is advanced incrementally as the packet bitmap is set and packets arrive. In the event of sporadic packet losses, which can leave a few bits in the bitmap set to 0, a time-out mechanism is used to advance the base SN.
Regarding responses to egress requests, the Trans-RX module first extracts the RIFT index from the packet header. The sequence number from the corresponding RIFT entry is used for validating the received response packet. Upon successful validation, the RIFT entry is cleared, and the response payload is forwarded to the CXL subsystem to complete the host transaction.

\begin{figure}[!h]
    \centering
    \includegraphics[width=\columnwidth,page=15]{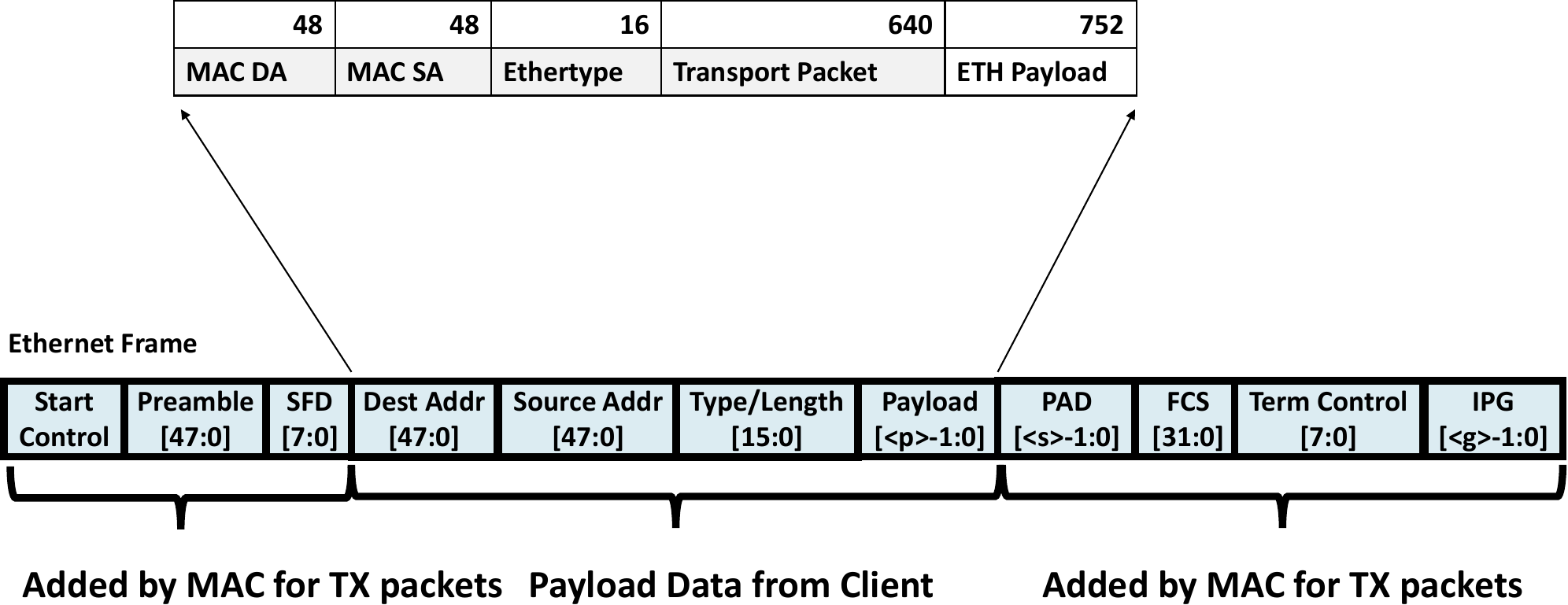}
    \caption{Ethernet frame format}
    \label{fig:11}  
\end{figure}

The Ethernet layer is composed of ETH-TX and ETH-RX modules, as shown in Figure \ref{fig:7}. The ETH-TX module receives transport packets from the Trans-TX module and constructs IEEE 802.3 Ether II frames.  Similarly, this module accepts control packets from ARM firmware and encapsulates them into Ethernet frames for transmission. The Ethernet frame format is shown in Figure \ref{fig:11}. The destination MAC address is obtained from the target device ID via a lookup in the MAC table. The firmware programs the source MAC address during the discovery phase. The Ethernet subsystem uses special Ethertype values for differentiating datapath and control packets. The fully formed Ethernet frames are then forwarded to the MAC for transmission via the MAC Interface Controller (MIC) module.

\begin{figure}[!h]
    \centering
    \includegraphics[width=\columnwidth,page=14]{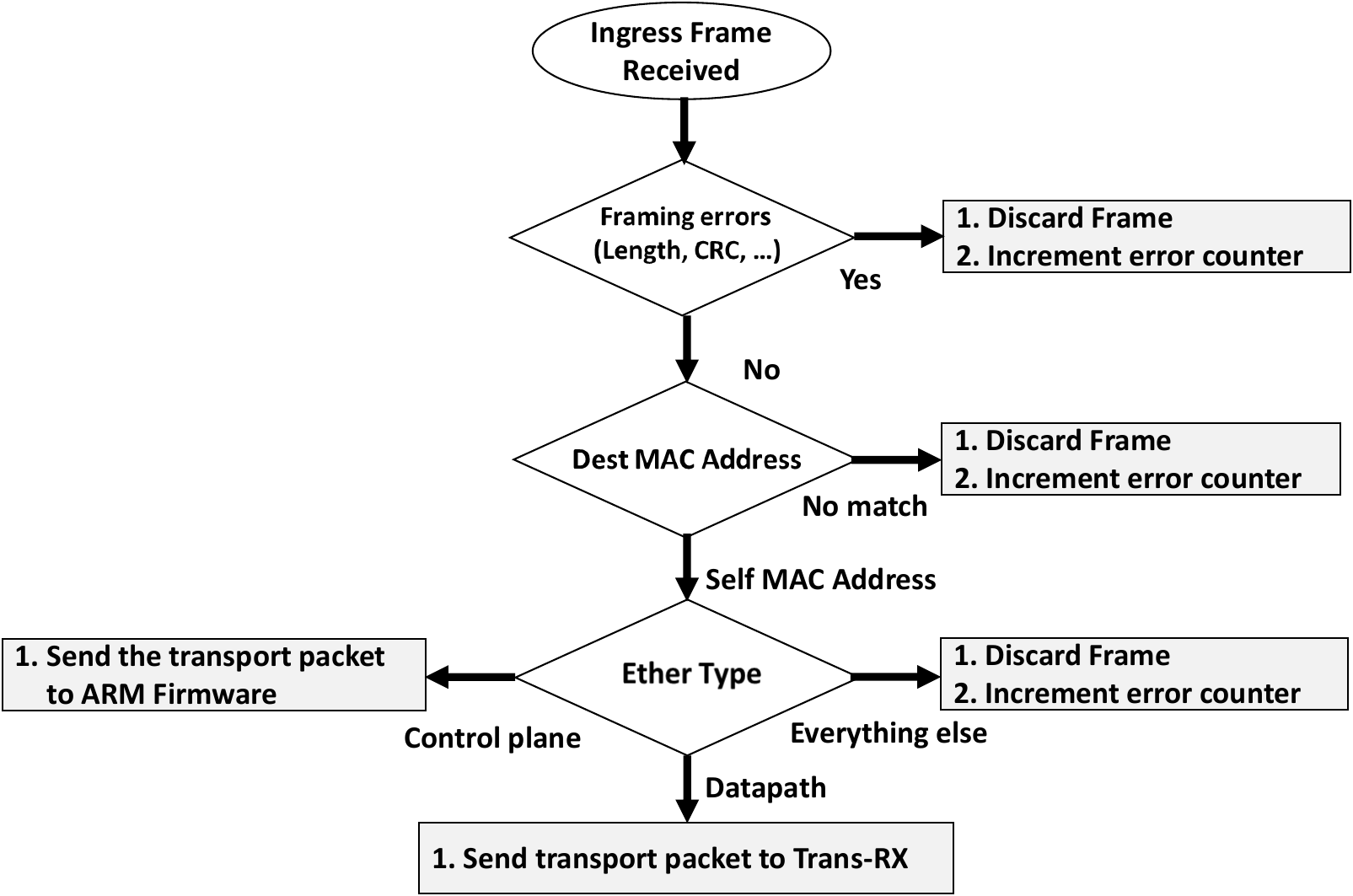}
    \caption{Algorithm for Ethernet frame processing}
    \label{fig:12}  
\end{figure}

The ETH-RX module processes the ingress Ethernet frames as shown in Figure \ref{fig:12}.

The MIC module implements the logic necessary to exchange Ethernet frames with MAC IP. It also performs clock-domain crossing (CDC) to isolate the MAC clock and the core clock domains. Ethernet MAC, PHY, and SERDES are hard macros used to implement the 802.3 Ethernet standard\cite{ref34}.    

\subsection{DDR Subsystem}
\begin{figure}[!h]
    \centering
    \includegraphics[width=\columnwidth,page=16]{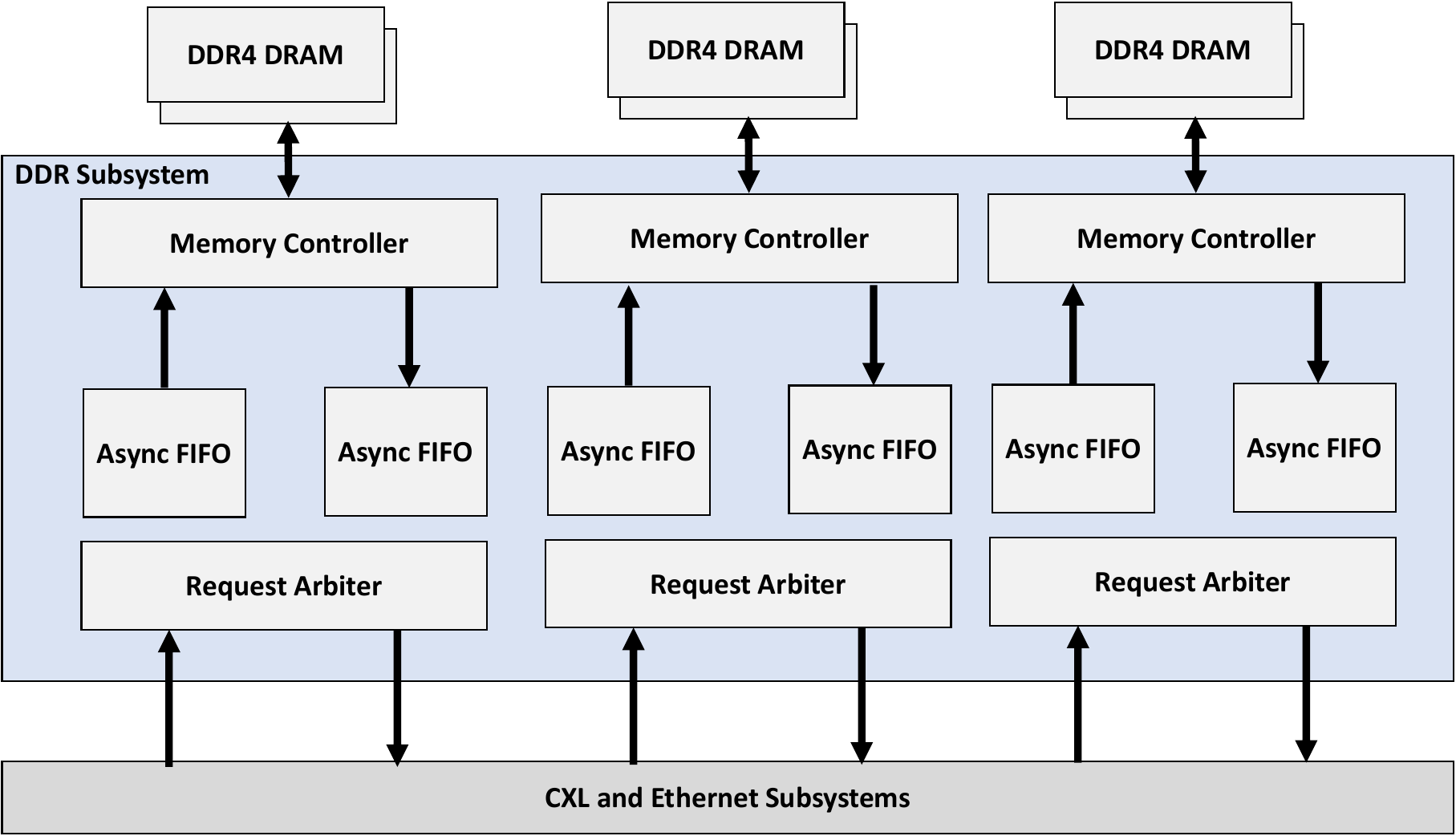}
    \caption{DDR Subsystem controller}
    \label{fig:13}  
\end{figure}

The DDR subsystem is a high-performance interconnect module that provides the access interface to actual DDR memory media. It supports three DDR4 channels, each with 16 GB of DRAM, delivering high-bandwidth, low-latency data access. As shown in Figure \ref{fig:13}, the DDR subsystem consists of a DDR4 memory controller, Clock Domain Crossing (CDC) logic using an asynchronous FIFO, and requester interfaces. The subsystem arbitrates concurrent read and write requests from multiple requesters from the CXL (local host accesses) and Ethernet (remote host accesses) subsystems. The DDR Subsystem stripes the memory addresses across its channels to maximize throughput while maintaining a strict request-response association. This mechanism ensures the data read from DDR memory is delivered to the correct requester. 

\subsection{ARM Subsystem}
\begin{figure}[!h]
    \centering
    \includegraphics[width=\columnwidth,page=8]{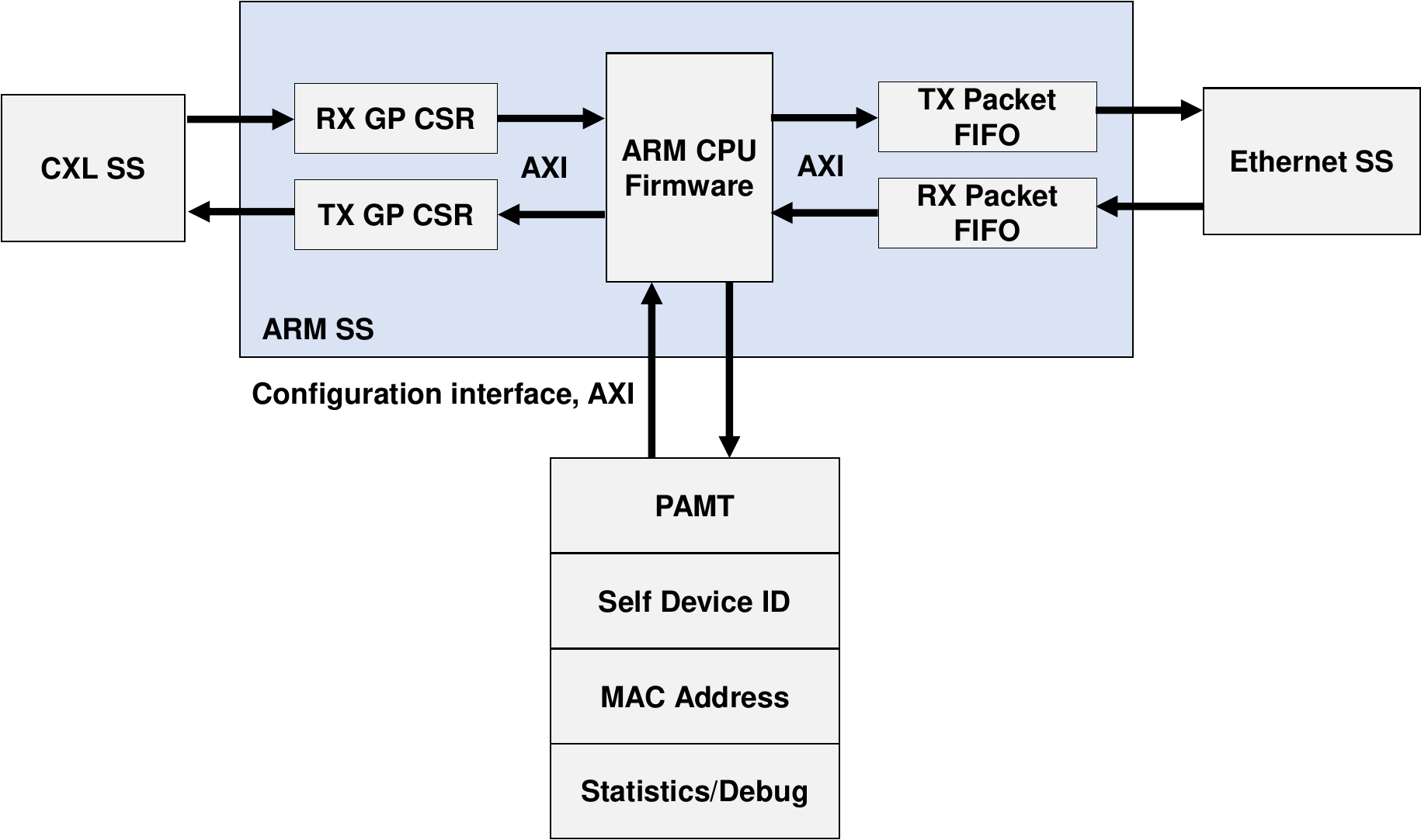}
    \caption{ARM subsystem micro architecture}
    \label{fig:14}  
\end{figure}

The ARM subsystem functions as the control plane of the disaggregated memory pool. CMM-MP firmware performs autonomous pool device discovery, pool memory map composition, and datapath configuration without host intervention. This subsystem comprises an ARM CPU core, a CXL General-Purpose Control and Status Register (GP-CSR) interface, and an Ethernet Packet interface module, as shown in Figure \ref{fig:14}. The Host software utility communicates with the ARM firmware via GP-CSR (Receive RX, Transmit TX). The host software sends commands and hints to the device by writing them into the RX GP-CSRs using the CXL.io protocol. To receive communication from the Host software, ARM firmware polls the RX GP-CSRs. The ARM firmware sends back responses and event information to the Host software via TX GP-CSRs. ARM firmware sends and receives Ethernet frames via the Transmit and Receive packet FIFOs, which interface with the datapath. The ARM subsystem uses a special Ethertype value in IEEE 802.3 Ethernet frames to distinguish control packets from data packets. After the successful discovery phase, ARM firmware configures the datapath by programming PAMT entries, device IDs, and MAC addresses. 
\subsection{CMM-MP Pool Device Discovery}
\begin{figure}[!h]
    \centering
    \includegraphics[width=\columnwidth,page=9]{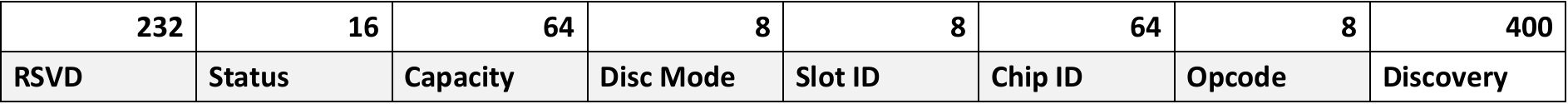}
    \caption{Discovery frame format}
    \label{fig:15}  
\end{figure}

One of the critical functions of ARM firmware is pool discovery. Discovery is the process of identifying and enumerating all available devices in the memory pool. During pool initialization, discovery is performed autonomously by device firmware running in the ARM subsystem. ARM firmware uses a discovery packet as shown in Figure \ref{fig:15}. After power-on, each device broadcasts a discovery message and listens for incoming messages. It compiles a sorted list of all unique responses, ordered by chip identifier. The rank of the chip identifiers in the sorted list is used to assign device IDs to the devices. These device IDs and the associated capacities are used to program the global pool map in PAMT. Upon successful completion of the discovery process, which runs concurrently across all devices, the memory pool transitions to the runtime state. 

\section{Host Interface}
This section describes the software interface and the memory allocation process for CMM-MP. CMM-MP advertises the pool memory as CXL Type 3, NUMA memory to the Host. Hosts can use CMM-MP as a memory resource in pooling, sharing, or both modes simultaneously. In pooling mode, all or part of the pool memory can be allocated to one or more hosts for exclusive access. Multiple hosts can access their allocated memory concurrently from the pool. In the sharing mode, an allocated pool memory address range can be accessed concurrently by multiple Hosts.
When pool memory is used as shared memory, CMM-MP devices guarantee atomicity of host accesses, but not the order of execution of host requests. For this reason, data coherency must be handled by the upper-layer software in the relevant hosts, if necessary. Such shared memory can be used in common usage modes such as producer-consumer and writer-reader, with explicit out-of-band synchronization and communication between participating hosts.    

\begin{figure}[!h]
    \centering
    \includegraphics[width=\columnwidth,page=17]{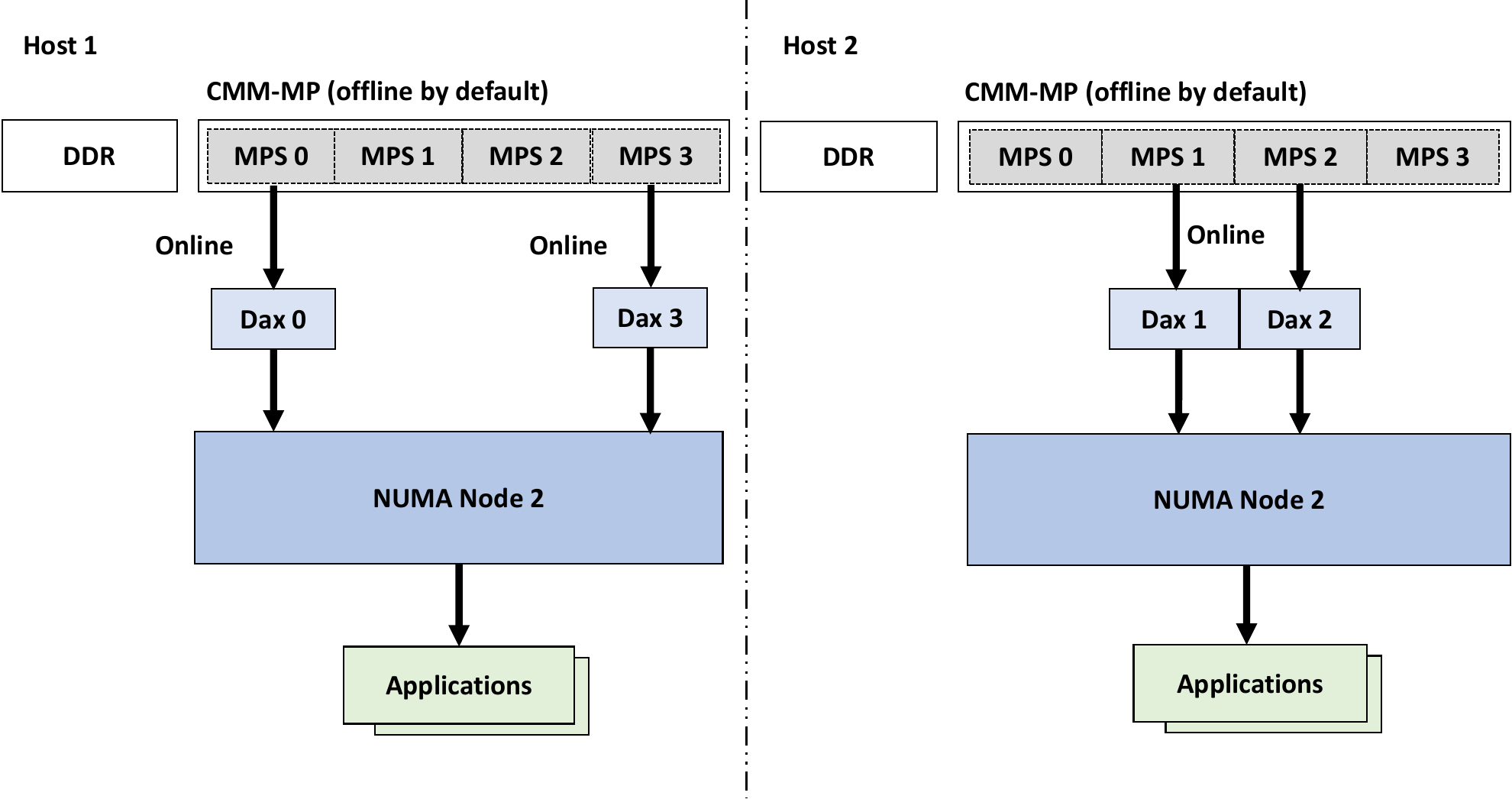}
    \caption{Memory allocation}
    \label{fig:16}  
\end{figure}

We developed a host management utility called Memory Pool Control (mpctl). It runs on each host and facilitates online or offline allocation of the memory pool segments. Each host reserves an address space equal to the entire memory pool using the Linux GRUB command. The memory pool region is divided into fixed, predefined memory segments using the memmap kernel parameter. All Memory Pool Segments (MPS) are offline by default. The Grub settings define the number and size of MPS that can be offlined/onlined.  
As shown in Figure \ref{fig:16}, the mpctl online MPS 0, 3 for Host 1 and MPS 1, 2 for Host 2. 

\section{Aggregated memory pool} 
\begin{figure}[!h]
    \centering
    \includegraphics[width=\columnwidth,page=18]{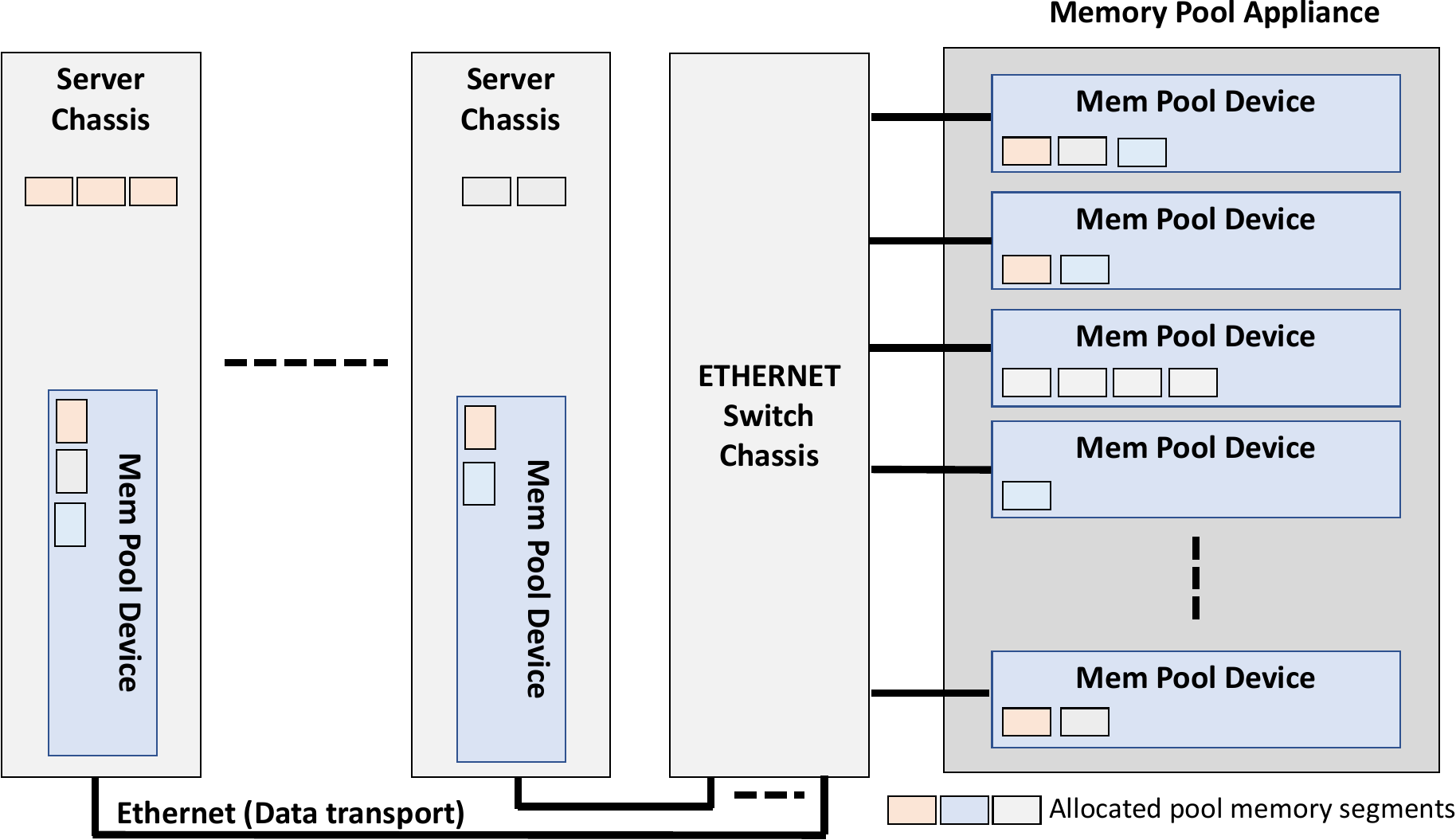}
    \caption{Aggregated Memory Pool}
    \label{fig:17}  
\end{figure}

\begin{figure}[!h]
    \centering
    \includegraphics[width=\columnwidth,page=20]{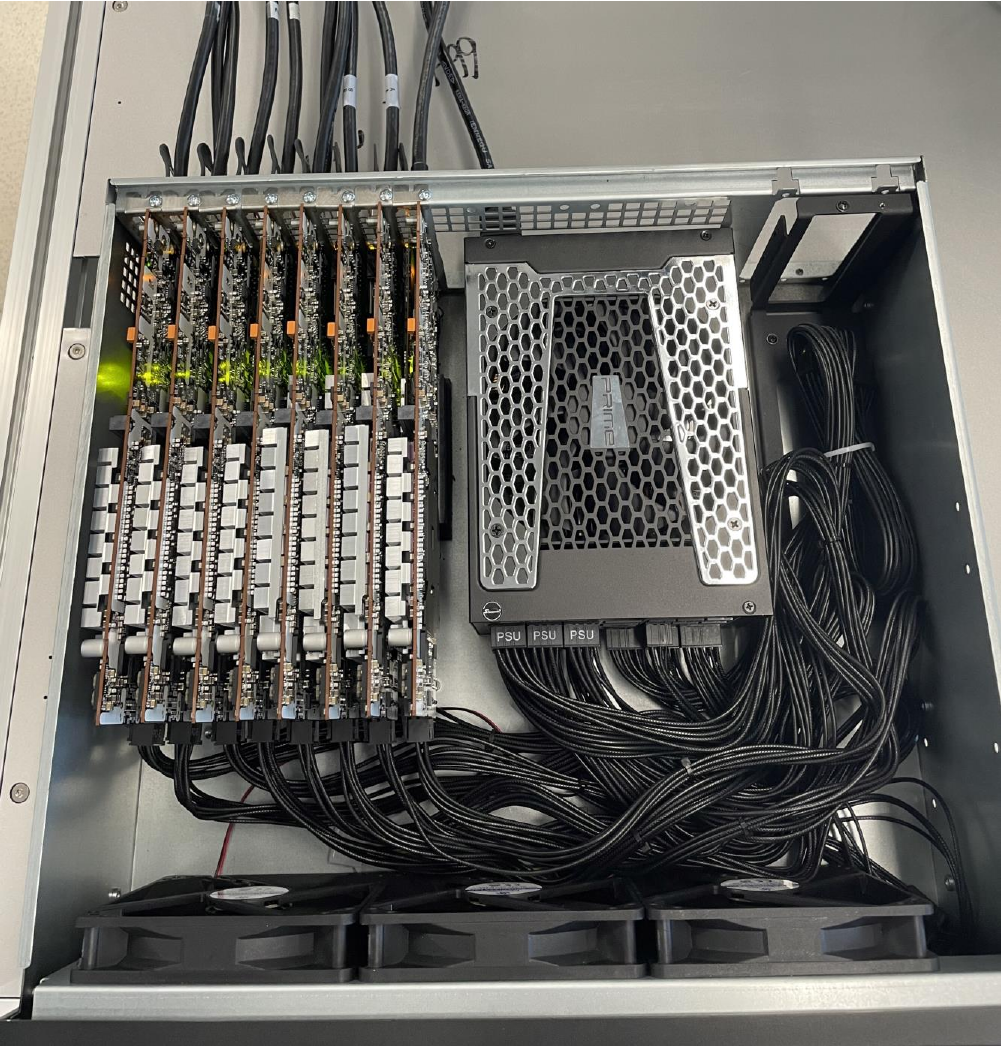}
    \caption{Example CMM-MP prototype system }
    \label{fig:20}  
\end{figure}

As shown in Figure \ref{fig:17}, CMM-MP devices can be used to create a scale-up memory pool solution that multiple hosts can share. The memory pool chassis contains the desired number of CMM-MP devices, which are connected through one or more dedicated Ethernet switches in the rack. Each host uses a local CMM-MP device to access the pool memory. The allocated memory segments (\textbf{\textit{small color boxes}}) can reside on different CMM-MP devices. In this scale-up system architecture, compute and memory can scale independently. Such system architecture enables potential use of different cooling methods (e.g., liquid or air cooling) for compute and memory. Figure \ref{fig:20} shows a prototype of an aggregated memory pool using 8 standalone CMM-MP devices.

\section{Distributed memory pool}
\begin{figure}[!h]
    \centering
    \includegraphics[width=\columnwidth,page=19]{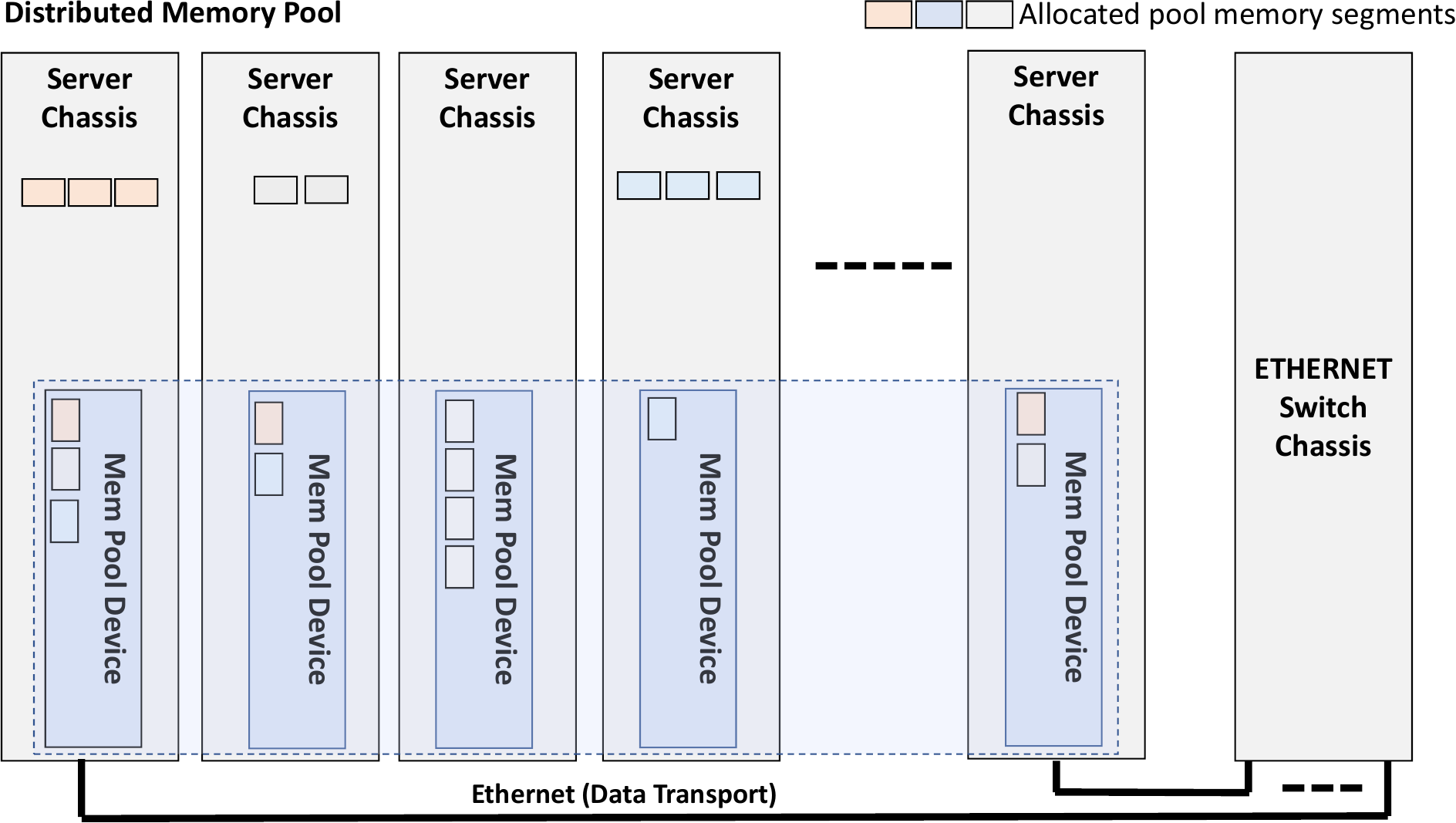}
    \caption{Distributed Memory Pool}
    \label{fig:18}  
\end{figure}

As an alternative to the aggregated memory pool solution described earlier, CMM-MP devices can also be used to create a distributed memory pool, as shown in Figure \ref{fig:18}. A set of hosts can pool together their CMM-MP device memories to create a large-capacity memory pool. The CMM-MP devices are connected using one or more dedicated Ethernet switches in the rack. Servers can be added or removed from the pool as needed. This system architecture provides the data center operator with the flexibility to use existing server resources more efficiently. Compared to the aggregated memory pool architecture, this solution tightly couples compute and memory for scaling while retaining the scalability advantage, as additional hosts can be added to the pool.  

\section{Performance Evaluation}
\subsection{Experiment Setup}
The evaluation is conducted on an Intel Granite Rapids (GNR)  development server with CXL support. The experimental setup consists of 2 host servers connected to a memory bank of 4 standalone CMM-MP devices. This creates a memory pool of 6 CMM-MP devices with a total capacity of 288 GB. The setup uses a 32-port Ethernet switch, each port capable of 200 Gbps, to connect all the CMM-MP devices. The detailed system configuration and CMM-MP specifications are provided in Table~\ref{table:system_config2}. 

\begin{table}[h!]
    \centering
    \caption{System Configuration-GNR}
    \resizebox{\columnwidth}{!}{
    \begin{tabular}{|l|l|}
    \hline
	\textbf{Platform}&\textbf{Component Description} \\        
	\hline
        \multirow{7}{*}{Server 1}
        & Intel (BeechnutCity) \\
        & Dual Socket Intel(R) Xeon(R) \\
        & 64 core per socket (Model 173) \\
        & System Memory: 512 GB DDR5-4800 MT/s \\  & System Memory Config: 8x32 GB per socket \\  
        & OS: Ubuntu 24.04 LTS (kernel 6.8.0-101) \\
        & NIC: Mellanox, [ConnectX-6] MT28908  \\
        & CMM-MP: Gen5x8, CXL1.1, 48 GB per device \\
        & SSD: Samsung PM1743, Gen5x4, 3.84 TB\\
	\hline
        \multirow{6}{*}{Server 2}
        & Intel (BeechnutCity) \\
        & Dual Socket Intel(R) Xeon(R)  \\
         & 144 core per socket (Model 175)\\
         &Memory: 256 GB DDR5-4800 MT/s \\    
         &OS: Ubuntu 24.04 LTS (kernel 6.8.0-101) \\
         &NIC: Mellanox, [ConnectX-6] MT28908 \\         
         &CMM-MP: Gen5x8, CXL1.1, 48 GB per device \\
    \hline
        \multirow{1}{*}{Standalone MP}
         &4x CMM-MP, 2x 200 Gbps ETH, 48 GB each\\
    \hline         
    \multirow{1}{*}{MP Switch}
        & NVIDIA MSN3700-VS2F 200G, 32x ports\\         
    \hline
	     \multirow{1}{*}{Network Switch}
         & NVIDIA MSN3700-VS2F 200G, MTU 9100  \\          
     \hline

    \end{tabular}
    }
    \label{table:system_config2}
\end{table}

\subsection{Device Characterization}
For device characterization, Intel MLC\cite{ref54} is employed.
First setup configuration is a direct connection between two nodes, i.e., no-switch (NS). The second configuration uses a switch (WS) to connect the two CMM-MP devices in the servers. The Intel MLC results for memory in local CMM-MP (MP-L), remote CMM-MP (MP-R) with no-switch (NS), and remote with switch (WS) on a GNR server are provided in Table~\ref{table:mlc_result_gnr_latency} and ~\ref{table:mlc_result_gnr_bw}. One observation from the latency table is that CMM-MP local memory latency is close to the cross-socket latency on the GNR server. Observing the latency data indicates that the Ethernet switch used in the experiments contributes a port-to-port round-trip delay of about 800 ns.

It is worth noting that in Table~\ref{table:mlc_result_gnr_bw} WS maintains highly consistent bandwidth across different read/write mixes, indicating a steady pipeline for remote memory accesses. For measurements in Table~\ref{table:mlc_result_gnr_bw}, one device is used for MP-L and five remote devices are used for MP-R (WS). 

\begin{table}[h!]
    \centering
    \caption{CMM-MP latency measurement}
    \begin{threeparttable}
	   \begin{tabular}{lccccccccc}
        \hline
         &&&&\multicolumn{2}{c}{\textbf{MP-R}} \\          
        \hline        \textbf{Numa}&\textbf{0}&\textbf{1}&\textbf{MP-L}&\textbf{NS}&\textbf{WS} \\        
        \hline
        \multirow{1}{*}{0} & \multicolumn{1}{l}{132} & \multicolumn{1}{l}{437}& \multicolumn{1}{l}{530}& \multicolumn{1}{l}{1911}& \multicolumn{1}{l}{2725}\\				
        \hline
        \multirow{1}{*}{1} & \multicolumn{1}{l}{428} & \multicolumn{1}{l}{128}& \multicolumn{1}{l}{869}& \multicolumn{1}{l}{2195}& \multicolumn{1}{l}{3070}\\
        \hline
     \end{tabular}
     % \begin{tablenotes}     
     % \end{tablenotes}
    \end{threeparttable}        
    \label{table:mlc_result_gnr_latency}
\end{table}

\begin{table}[h!]
    \centering
    \caption{CMM-MP bandwidth measurement}
    \begin{threeparttable}
	   \begin{tabular}{llllll}
        \hline
        \textbf{Max BW (GB/s)}&\textbf{MP-L}&\textbf{MP-R~(WS, 5x CMM-MP)} \\        
        \hline
        \multirow{1}{*}{All Reads} & \multicolumn{1}{c}{26.8} & \multicolumn{1}{c}{23.1}\\				
        \hline
        \multirow{1}{*}{3:1 RW} & \multicolumn{1}{c}{17.5} & \multicolumn{1}{c}{24.4}\\			
        \hline
        \multirow{1}{*}{2:1 RW} & \multicolumn{1}{c}{19.9} & \multicolumn{1}{c}{24.3}\\			        
        \hline
        \multirow{1}{*}{1:1 RW} & \multicolumn{1}{c}{16.4} & \multicolumn{1}{c}{25.2}\\			           
        \hline
        \multirow{1}{*}{Stream-triad} & \multicolumn{1}{c}{22.2} & \multicolumn{1}{c}{23.8}\\			                   
        \hline
     \end{tabular}
     % \begin{tablenotes}
     % \end{tablenotes}
    \end{threeparttable}        
    \label{table:mlc_result_gnr_bw}
\end{table}

For comparison with traditional disaggregated memory solutions, we also measure RDMA performance on this experimental setup. The RDMA latency and bandwidth are shown in Table~\ref{table:rmda}.
For small transfers (64 B), RDMA operations exhibit microsecond-level latency, whereas large transfers (8 MB) achieve approximately 23 GB/s bandwidth. These measurements provide a baseline for evaluating remote memory access over conventional RDMA-based architectures. The application-level performance of remote memory access via RDMA is affected by additional overhead from process context switching and scheduling. Whereas the MLC benchmark is a traditional memory characterization benchmark.

\begin{table}[h!]
    \centering
    \caption{RDMA performance measurement}
	\begin{tabular}{lcccc}
        \hline
        	\textbf{RDMA metric}&\textbf{Type}&\textbf{Size}&\textbf{NS}&\textbf{WS} \\        
        \hline
        \multirow{3}{*}{Latency (us)} & 
        \multicolumn{1}{l}{read} & 
        \multicolumn{1}{c}{64B} & \multicolumn{1}{c}{2.27} & \multicolumn{1}{c}{3.17}\\		
        \cline{2-5}
        &\multicolumn{1}{l}{write} & 
        \multicolumn{1}{c}{64B} & \multicolumn{1}{c}{1.36} & \multicolumn{1}{c}{1.77}\\		
        \cline{2-5}        
        &\multicolumn{1}{l}{send} & 
        \multicolumn{1}{c}{64B} & \multicolumn{1}{c}{1.48} & \multicolumn{1}{c}{1.85}\\		
        \hline
        \multirow{6}{*}{BW (GB/s)} & 
        \multicolumn{1}{l}{read} & 
        \multicolumn{1}{c}{64B} & \multicolumn{1}{c}{0.16} & \multicolumn{1}{c}{0.15}\\		
        \cline{2-5}
        &\multicolumn{1}{l}{write} & 
        \multicolumn{1}{c}{64B} & \multicolumn{1}{c}{0.23} & \multicolumn{1}{c}{0.19}\\		
        \cline{2-5}        
        &\multicolumn{1}{l}{send} & 
        \multicolumn{1}{c}{64B} & \multicolumn{1}{c}{0.16} & \multicolumn{1}{c}{0.15}\\	
        \cline{2-5}        
        &\multicolumn{1}{l}{read} & 
        \multicolumn{1}{c}{8M} & \multicolumn{1}{c}{23} & \multicolumn{1}{c}{23}\\	\cline{2-5}        
        &\multicolumn{1}{l}{write} & 
        \multicolumn{1}{c}{8M} & \multicolumn{1}{c}{24} & \multicolumn{1}{c}{24}\\	
        \cline{2-5}        
        &\multicolumn{1}{l}{send} & 
        \multicolumn{1}{c}{8M} & \multicolumn{1}{c}{24} & \multicolumn{1}{c}{24}\\	
        \hline

    \end{tabular}
    \label{table:rmda}
\end{table}

\begin{figure}[!h]
    \centering
    \includegraphics[width=\columnwidth,page=28]{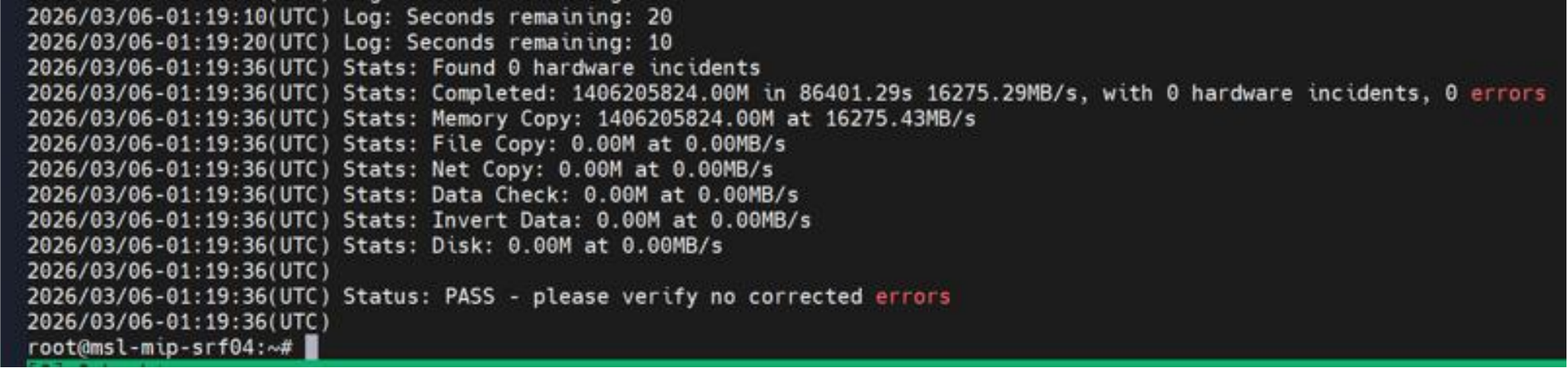}
    \caption{GSAT report}
    \label{fig:28}  
\end{figure}

To verify the stability of the CMM-MP system, we run the Google Stressful Application Test (GSAT)\cite{ref57}, a widely used memory stress-testing tool. GSAT continuously exercises the memory subsystem and detects potential memory errors during long-running execution. Figure \ref{fig:28} shows the successful completion of a 24-hour run with no detected errors.

\subsection{Real World Application Use Cases}

To evaluate the performance of memory pooling on real-world database workloads, Valkey\cite{ref58} database, an in-memory key–value store, is employed. We performed Valkey native micro-benchmark (Valkey-benchmark) as well as a more application-oriented benchmark, YCSB\cite{ref55}.

\begin{figure}[!h]
    \centering
    \includegraphics[width=\columnwidth,page=24]{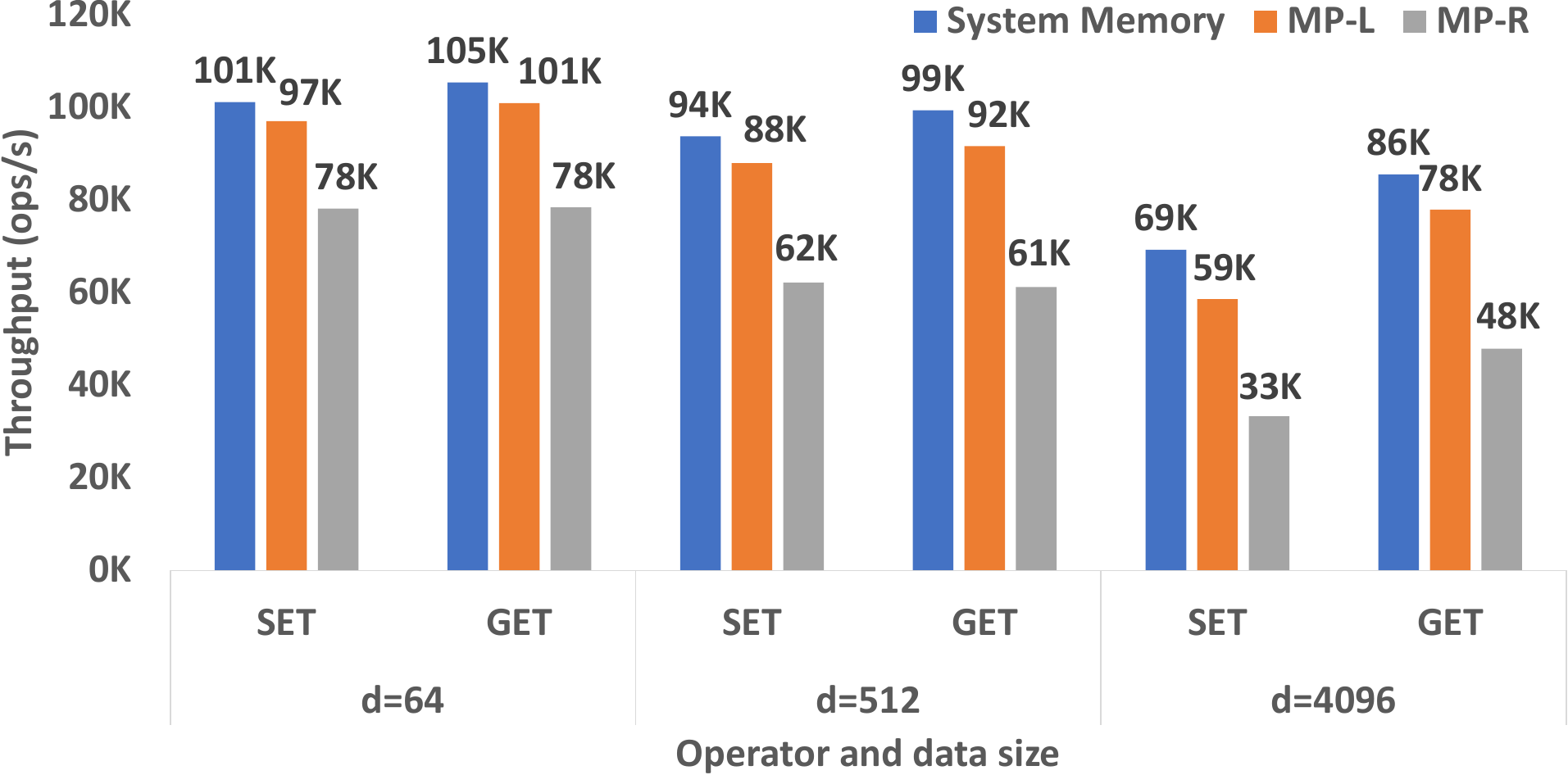}
    \caption{Valkey benchmark throughput}
    \label{fig:23}  
\end{figure}

Figure \ref{fig:23} shows the Valkey benchmark throughput for SET and GET operations while varying the value size from 64 up to 4096 bytes. As the value size increases, throughput decreases across all configurations due to the larger data transfer volume. MP-L remains close to the system memory baseline across all value sizes. MP-R shows lower throughput compared with MP-L because larger values require more data to be transferred over the backend Ethernet network.
As seen in the figure \ref{fig:23}, even though CMM-MP has 10x less raw bandwidth compared to the system memory (32 GB/s vs 304 GB/s), the Valkey throughput performance degradation is below 50\% for these tests. Note that system memory bandwidth is calculated using 38 GB/s DDR5 4800 MT/s single-channel bandwidth and 8 channels per socket in the test server.     

\begin{figure}[!h]
    \centering
    \includegraphics[width=\columnwidth,page=25]{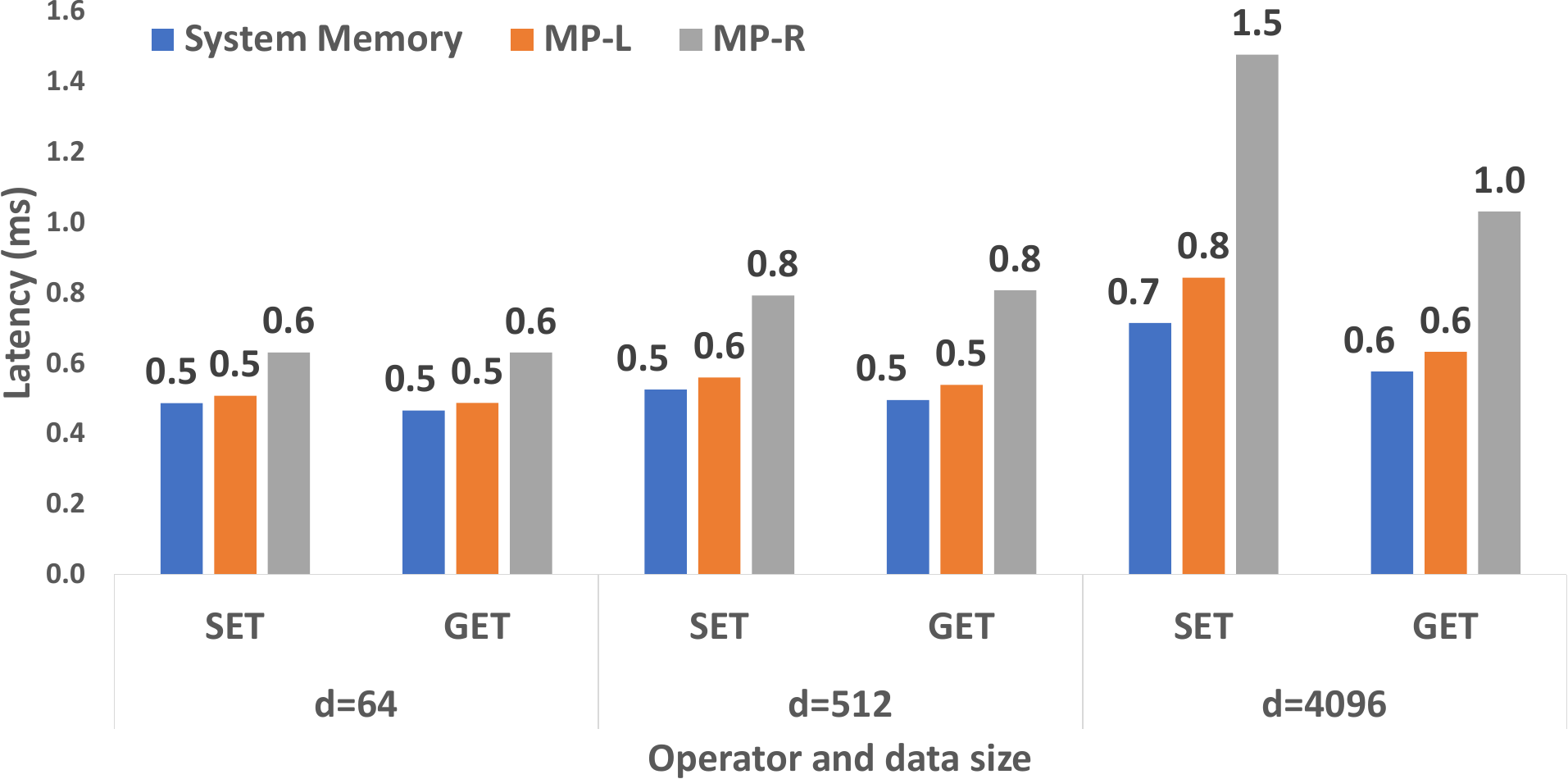}
    \caption{Valkey benchmark latency}
    \label{fig:24}  
\end{figure}

Figure \ref{fig:24} reports the corresponding average latency for Valkey. For both SET and GET operations, MP-L latency remains close to system memory across different value sizes, demonstrating that local pooled memory behaves similarly to local System memory. MP-R shows higher latency due to the additional network transport required for remote memory accesses. When the dataset size is small, the latency performance degradation remains moderate around 20\%.  

In addition to the Valkey benchmark, Valkey is used alongside the Yahoo Cloud Serving Benchmark (YCSB). YCSB provides multiple workload patterns representing different read/write ratios.

\begin{figure}[!h]
    \centering
    \includegraphics[width=\columnwidth,page=22]{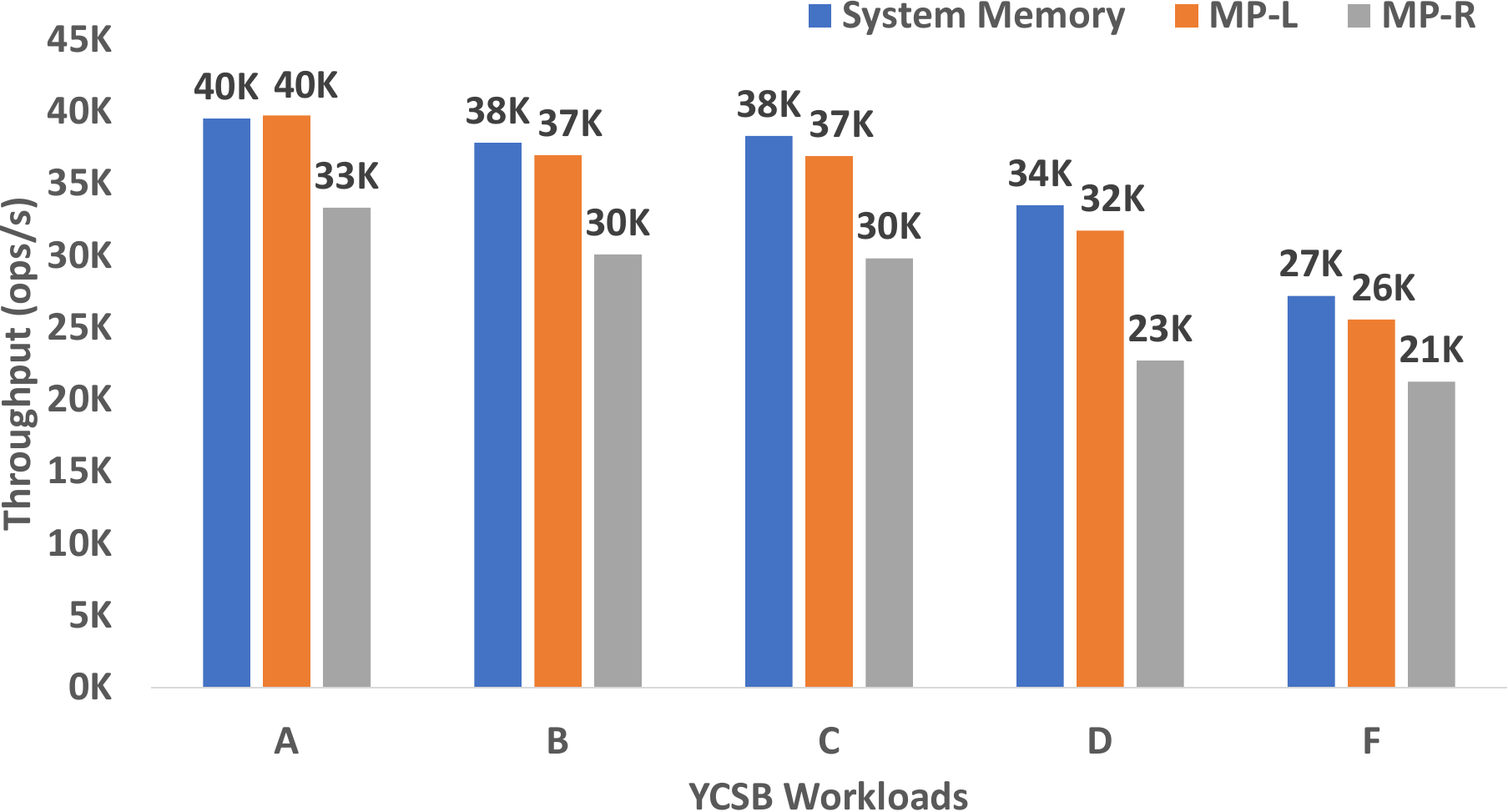}
    \caption{Valkey+YCSB throughput}
    \label{fig:21}  
\end{figure}

Figure \ref{fig:21} shows the throughput for YCSB workloads A-F across three memory configurations: System memory, local pooled memory (MP-L), and remote pooled memory (WS). Across all workloads, MP-L achieves performance comparable to system memory, indicating that the CXL-attached memory pool incurs minimal overhead for local accesses. Remote pooled memory exhibits lower throughput due to the additional network hop, but still maintains substantial performance. 

As seen in the figure \ref{fig:21}, even though CMM-MP has 10x less raw bandwidth compared to the system memory (32 GB/s vs 304 GB/s), the Valkey + YCSB throughput performance degradation is well below 30\% for these tests, indicating CMM-MP's potential for TCO optimization.  

\begin{figure}[!h]
    \centering
    \includegraphics[width=\columnwidth,page=23]{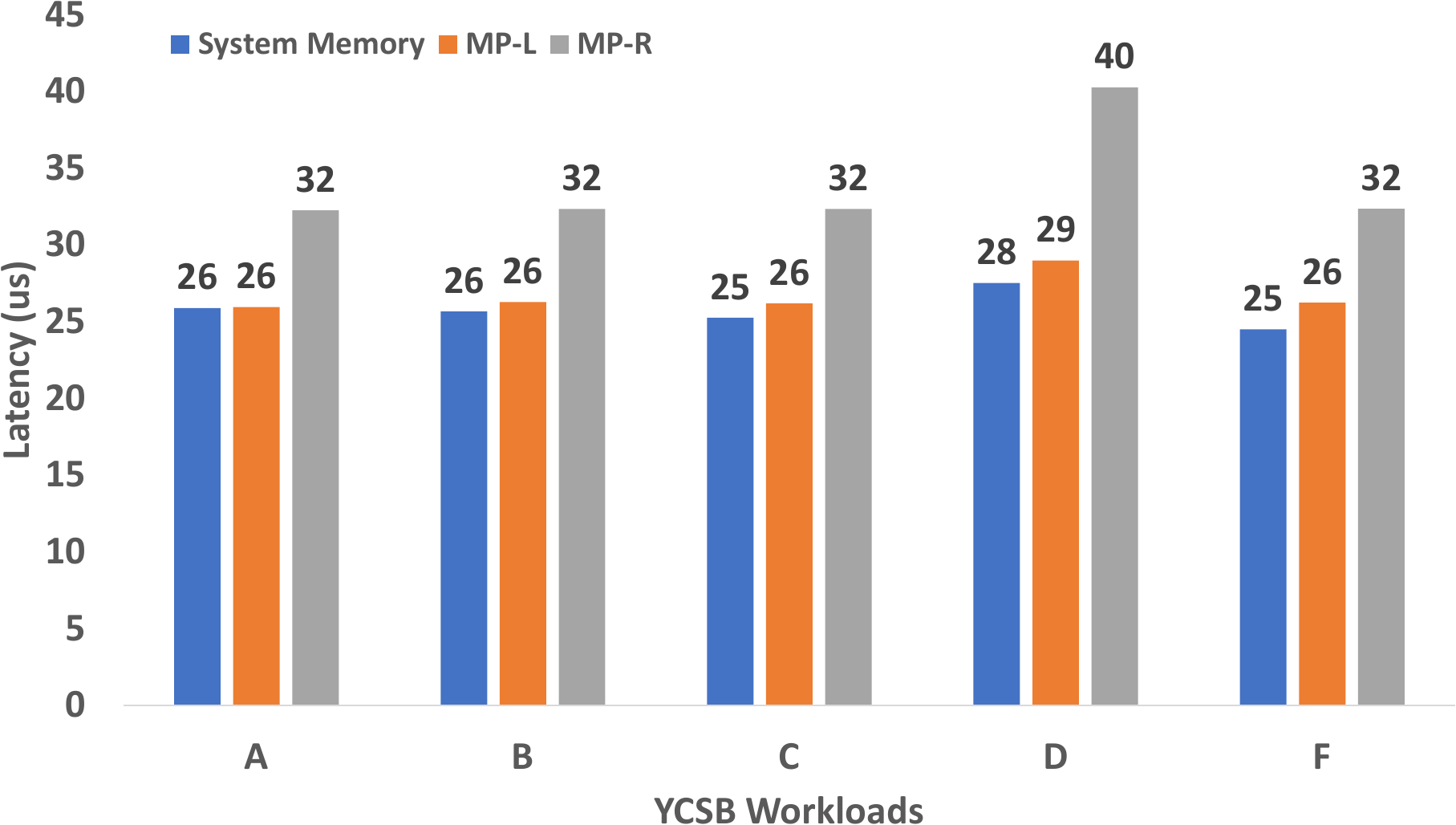}
    \caption{Valkey+YCSB Latency}
    \label{fig:22}  
\end{figure}

Figure \ref{fig:22} shows the corresponding Valkey YCSB latency results. MP-L exhibits latency comparable to system memory across all workloads, whereas MP-R incurs higher latency due to backend Ethernet communication, but the increase remains moderate and consistent across all workloads. The latency degradation remains about 30\% for all workloads, except D. Workload D has a 5\% insert rate, which causes greater network write overhead.     

\begin{figure}[!h]
    \centering
    \includegraphics[width=\columnwidth,page=26]{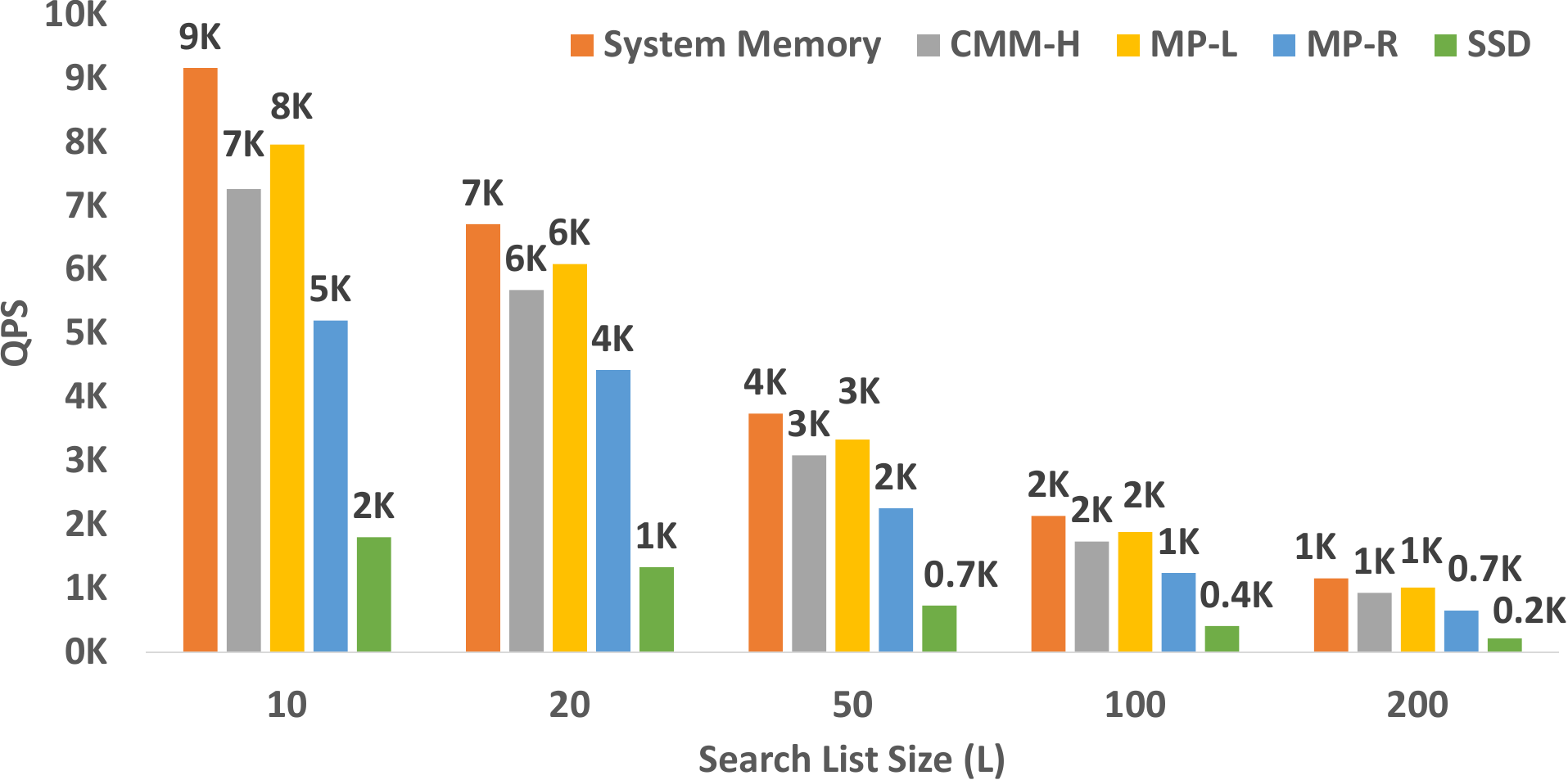}
    \caption{DiskANN}
    \label{fig:26}  
\end{figure}

Another use case is DiskANN\cite{ref56}, a high-performance approximate nearest-neighbor (ANN) search system designed for very large graph vector datasets. In this evaluation, we use the SIFT dataset with 1 million records, and one single CPU core to run the Search\_disk\_index algorithm. For all the experiments, memory is exposed as a ramdisk to ensure a fair comparison. Figure \ref{fig:26} shows the query throughput (QPS) of DiskANN under different search list sizes for five configurations: system memory, CMM-H, MP-L, MP-R, and SSD. System memory achieves the highest throughput, while MP-L closely follows the System memory baseline across all search sizes (L), indicating that the CMM-MP architecture can effectively support memory-intensive vector search workloads. MP-R exhibits lower throughput due to the cost of remote memory access over the backend network, yet it still significantly outperforms disk-based storage by a factor of about 3x. These results demonstrate that pooled memory can offer substantially higher performance than disk for large-scale vector search workloads.

\section{Conclusion}
\label{sec:conclusions}
In this paper, we propose CMM-MP, a highly cost-effective yet scalable memory pooling architecture. To the best of our knowledge, we have developed an industry-first, purpose-built memory pooling device that enables flexible and scalable compute and memory disaggregation. CMM-MP requires no application changes to utilize the memory pool and preserves traditional software interfaces such as load/store, devdax, and fsdax. This architecture offloads and abstracts all the memory pool composition and management functions from the Host OS to the device. The CMM-MP device effectively uses Ethernet to connect multiple memory devices on the backend while presenting a standard CXL Type 3 memory interface to the host. Our experiments demonstrate that CMM-MP can provide an easy-to-use solution compared to traditional RDMA-based solutions, while delivering significantly lower latency and higher throughput. CMM-MP provides a practical, cost-effective foundation for future memory-centric data centers.  
Our future work will focus on developing more advanced features. We plan to support emerging scale-up and scale-out backend protocols, such as Ethernet Scale Up Network (ESUN), Scale Up Ethernet - Transport (SUE-T), Ultra Accelerator Link (UALink), and Ultra Ethernet (UE), for efficient, high-bandwidth data transfers to xPUs. We also plan to explore device-side caching to fully capture and utilize application workload spatial and temporal localities. Another focus area is to extend current static memory management to a dynamic approach for more robust pool usage. We also plan to explore and investigate fine granular, dynamic allocation and management of pool memory. 

\section*{Acknowledgment}
The authors thank Dr. Sungwook Ryu for his guidance and support. We also express our gratitude to all members of the MSL DCDS team for their technical contributions to the CMM-MP development. We would like to thank CVP Hojun Shim and CVP Taeksang Song for their valuable suggestions and review comments. 

\bibliographystyle{IEEEtran}
\bibliography{src/references}

\begin{IEEEbiographynophoto}{Michael Wasef}
Michael is a senior RTL design engineer in the Memory Solutions Lab at Samsung, focused on developing new, innovative memory device solutions. Received his PhD from Boise State University. He specialized in device architectures for next-generation memory ecosystems.
His work primarily focuses on CXL (Compute Express Link) technologies, specifically in the areas of memory pooling, software-defined memory. Wasef plays a key role in bridging the gap between cutting-edge hardware and system-level software, ensuring high-performance memory solutions are optimized for AI, machine learning, and modern data center infrastructures.
\end{IEEEbiographynophoto}
\vspace*{-2\baselineskip}

\begin{IEEEbiographynophoto}{Gongjin Sun} is a Senior Architect with the Memory Solutions Lab at Samsung Semiconductor, Inc. His work focuses on the design and optimization of next-generation memory-centric computing systems, including CMM-H and CMM-MP. He received the Ph.D. degree from the University of California, Irvine. His research spans computer architecture, CXL-based memory and storage, memory offloading for large language model (LLM) inference, FPGA-based acceleration, near-storage acceleration, Linux kernel development, and data compression.
\end{IEEEbiographynophoto}
\vspace*{-2\baselineskip}

\begin{IEEEbiographynophoto}{Edmund Au}  is a staff design engineer in Memory Solutions Lab at Samsung, With over 25 years of semiconductor experience, he is a key figure in architecting next-generation storage and memory systems.
He focuses on the development of CXL (Compute Express Link) controllers, and advanced SSD subsystems. 
\end{IEEEbiographynophoto}
\vspace*{-2\baselineskip}

\begin{IEEEbiographynophoto}{Ranjitha Gopalakrishna}  is a senior verification engineer in Memory Solutions Lab at Samsung. specialist in test environment architecture, she focuses on the development and optimization of CXL (Compute Express Link) technologies and next-generation memory controllers.

\end{IEEEbiographynophoto}
\vspace*{-2\baselineskip}

\begin{IEEEbiographynophoto}{Elham Movahedian}  is a staff engineer in Memory Solutions Lab at Samsung, A specialist in computer architecture and system performance, she is a key contributor in the development of next-generation memory systems.
\end{IEEEbiographynophoto}
\vspace*{-2\baselineskip}

\begin{IEEEbiographynophoto}{Dongyang Li}  is a senior RTL design engineer in Memory Solutions Lab at Samsung. Working in the research and development of advanced memory architectures. He is a key contributor to the evolution of CXL (Compute Express Link) technologies, with a focus on memory pooling and fabric management for data centers.
His work is instrumental in optimizing low-latency memory solutions for AI-driven workloads and large-scale distributed systems.
\end{IEEEbiographynophoto}
\vspace*{-2\baselineskip}

\begin{IEEEbiographynophoto}{Senthil Murugesapandian}  is a staff engineer in Memory Solutions Lab at Samsung, in San Jose, California. His work primarily focuses on advancing CXL (Compute Express Link) ecosystems, SSD controller performance, and hardware-software co-design. 
\end{IEEEbiographynophoto}
\vspace*{-2\baselineskip}

\begin{IEEEbiographynophoto}{Caroline Kahn}  is a Principal Engineer working on firmware and software development. She contributes to the advancement of high-performance storage ecosystems. Her work ensures seamless collaboration across multi-disciplinary teams, driving the successful delivery of innovative memory and storage solutions that power modern data centers and cloud infrastructure.
\end{IEEEbiographynophoto}
\vspace*{-2\baselineskip}

\begin{IEEEbiographynophoto}{Shuyu Lyu}  is a staff RTL Design Engineer in Memory Solutions Lab at Samsung. Her work is central to the development of CXL (Compute Express Link) ecosystems, where she focuses on memory pooling, fabric management, and hardware-software co-optimization.
With a background in computer engineering, Lyu plays a key role in designing scalable, high-performance memory solutions. 
\end{IEEEbiographynophoto}
\vspace*{-2\baselineskip}

\begin{IEEEbiographynophoto}{Brian Luu}  is a staff verification engineer in Memory Solutions Lab at Samsung. His work focuses on integrating CXL (Compute Express Link) with high-performance NVMe subsystems. Luu role in system-level Verification, ensuring the performance of next-generation memory modules in complex data center environments. His expertise in hardware-software interaction is vital for optimizing low-latency solutions tailored for AI infrastructure and large-scale cloud computing.
\end{IEEEbiographynophoto}
\vspace*{-2\baselineskip}

\begin{IEEEbiographynophoto}{Jason Scott}  is a staff engineer in Memory Solutions Lab at Samsung, working on performance engineering of CXL memory prototypes. He plays a key role in the research and development of CXL (Compute Express Link) and persistent memory technologies, specifically working on system-level integration and benchmarking.
\end{IEEEbiographynophoto}
\vspace*{-2\baselineskip}

\begin{IEEEbiographynophoto}{Namita Kumar}  is a staff verification engineer in Memory Solutions Lab at Samsung. specialist in test environment architecture, she focuses on the development and optimization of CXL (Compute Express Link) technologies and next-generation memory controllers.
\end{IEEEbiographynophoto}
\vspace*{-2\baselineskip}

\begin{IEEEbiographynophoto}{Mehdi Nik} received the Ph.D. degree from the University of Pittsburgh. He is currently a Senior Staff Software Engineer with Samsung Memory Solutions Lab (MSL), where he works on high-performance computing (HPC) and large-scale simulation frameworks for next-generation memory and data-center systems. His technical interests include performance engineering, system-level optimization, and the characterization and analysis of emerging memory and storage devices.
\end{IEEEbiographynophoto}

\begin{IEEEbiographynophoto}{Kwan Huen}  Hingkwan (H.K.) Huen is a principal engineer and architect at Samsung’s Memory Solutions Lab (MSL). Specializing in storage virtualization and software-defined storage, he plays a critical role in developing advanced architectures for NVMe-based systems and cloud-scale data centers. With a deep expertise in Linux kernel development and high-performance drivers, Huen focuses on optimizing the interface between host software and device hardware.
\end{IEEEbiographynophoto}
\vspace*{-2\baselineskip}

\begin{IEEEbiographynophoto}{Xuebin Yao}  Xuebin Yao is a researcher at Samsung’s Memory Solutions Lab (MSL), where he focuses on advancing the frontiers of memory and storage technologies. With expertise in distributed systems, high-performance computing, and Flash-based storage architectures, his work is pivotal in optimizing data center efficiency and system-level performance.
\end{IEEEbiographynophoto}
\vspace*{-2\baselineskip}

\begin{IEEEbiographynophoto}{Amir Beygi} received the Ph.D. degree from Western University, London, ON. He is currently a Senior Staff Engineer with Samsung Memory Solutions Lab (MSL), where he designs next-generation memory and data-center systems. His research interests include RTL design and architecture for memory- and storage-centric systems, as well as computational storage for data center and AI infrastructure, with an emphasis on improving performance, scalability, and efficiency in large-scale systems.
\end{IEEEbiographynophoto}
\vspace*{-2\baselineskip}

\begin{IEEEbiographynophoto}{Luis Vitorio Cargnini}  (IEEE) received the Ph.D. degree in Microelectronics/Spintronics from the University of Montpellier, LIRMM, France, in 2013, specializing in MRAM. He focuses in high-performance computing (HPC) and hardware security, SoC, Computer architecture, Network-on-Chip (NoC) design, and energy-efficient domain-specific accelerators. 
\end{IEEEbiographynophoto}
\vspace*{-2\baselineskip}

\begin{IEEEbiographynophoto}{Ramdas Kachare}  Ramdas Kachare is a senior director and senior principal engineer at Samsung’s Memory Solutions Lab (MSL), where he leads innovation in disaggregated memory device solutions, Solid State Drive (SSD) architecture and storage systems. His research includes NAND flash management, CXL, PCIe/NVMe protocols, and data center memory and storage optimization. Kachare holds numerous patents and has been a key contributor to industry standards. His leadership at Samsung MSL focuses on next generation memory and storage device hardware capabilities.
\end{IEEEbiographynophoto}
\vspace*{-2\baselineskip}
\vfill 

\end{document}